# Sequence-based protein-protein interaction prediction and its applications in drug discovery

François Charih[1,2,3], James R. Green[1,3,†], and Kyle K. Biggar[2,3,†]

[1] Department of Systems and Computer Engineering, Carleton University, Ottawa, ON, Canada

[2] Institute of Biochemistry, Department of Biology, Carleton University, Ottawa, ON, Canada

[2] NuvoBio Corp., Ottawa, ON, Canada

[†] Contributed equally

## Abstract

Aberrant protein-protein interactions (PPIs) underpin a plethora of human diseases, and disruption of these harmful interactions constitute a compelling treatment avenue. Advances in computational approaches to PPI prediction have closely followed progress in deep learning and natural language processing. In this review, we outline the state-of-the-art for sequence-based PPI prediction methods and explore their impact on target identification and drug discovery. We begin with an overview of commonly used training data sources and techniques used to curate these data to enhance the quality of the training set. Subsequently, we survey various PPI predictor types, including traditional similarity-based approaches, and deep learning-based approaches with a particular emphasis on the transformer architecture. Finally, we provide examples of PPI prediction in systems-level proteomics analyses, target identification, and design of therapeutic peptides and antibodies. We also take the opportunity to showcase the potential of PPI-aware drug discovery models in accelerating therapeutic development.

## Introduction

Understanding how proteins interact is central to deciphering the molecular machinery of life. Protein-protein interactions (PPIs)underlie virtually every cellular process, from signal transduction to immune surveillance, and are essential for maintaining homeostasis across organisms. As our ability to probe these interactions has grown, through both experimental and computational tools, so too has our appreciation for their complexity and functional importance. In particular, disruptions or aberrant formations of PPIs have emerged as key contributors to disease, transforming our view of PPIs from abstract molecular partnerships into tangible drug targets. With the explosion of proteomic data and advances in artificial intelligence, the field has entered a new phase where it is now possible to accurately predict PPIs at proteome scale. This review explores how sequence-based PPI prediction has evolved into a critical tool for both basic biology and drug discovery.

Indeed, PPIs are the most prevalent type of interaction involving biomacromolecules [1]. They can be quasi-permanent, transient, functionally obligate, or non-obligate [2]. Virtually all biological processes involve PPIs whereby proteins physically interact to exert their function in an elegant and concerted fashion: DNA repair [3] and transcription [4], protein translation [5], cell signaling [6], and protein quality control [7] to name a few. As previously highlighted, it should therefore come as no surprise that abnormal PPIs are the main culprit in a wide range of human diseases [8], [9], [10]. Certain diseases are caused by, or influenced by unwanted interactions. For example, this is notably the case in numerous neurodegenerative disorders, where protein aggregation in neural tissue is a key feature: Alzheimer's disease ($\beta$ amyloid and tau protein) [11], amyotrophic lateral sclerosis (TDP-43) [12], Parkinson's disease ($\alpha$-synuclein) [13], Huntington's disease (Huntingtin) [14], and Creutzfeldt-Jakob (prion protein) [15], to name a few. Another common way through which PPIs can cause disease is through changes in interaction affinity following a mutation, as is the case for mutations in KRAS which impact interactions with its effectors [16]. A well-known example of a KRAS mutation affecting protein-protein interaction affinity is the KRAS G12D mutation, which impacts its interaction with GTPase-activating proteins (GAPs) and downstream effectors like RAF kinases. Altered protein stoichiometry caused by disruption of normal gene expression patterns leads to alterations in the PPI network and is also another key component of cancer [17], [18], [19].

The human proteome is currently believed to contain about 20,000 proteins – excluding isoforms resulting from alternative splicing or proteins modified post-translationally. As such, the total number of possible pairwise protein interactions is likely on the order of *at least* 200 million. Proteins selectively interact with a fraction of all possible interaction partners; though, how many of these protein pairs physically interact in a biologically relevant context is still open to speculation. To help validate PPIs, a variety of biochemical and biophysical techniques to detect interactions are commonly used and include yeast two-hybrid, affinity-purification coupled with mass spectrometry, phage display, pull-down assays, to name a few. Interested readers may refer to the following review articles for details regarding these experimental methods [20], [21], [22], [23], [24].

Because experimental techniques are resource-intensive, expensive, and are limited in their throughput, scientists increasingly rely on *in silico* predictors to identify which potential interactions they should prioritize for *in vitro* investigations and validation experiments. The earliest families of PPI predictors, surveyed in [25], largely relied, in explicit ways, on genomic (co-localization of genes), evolutionary (sequence co-evolution), and structural information (presence of binding motifs and domains). These approaches have largely been superseded by machine learning (ML)-based models, though some of these older approaches remain in use [26], [27], [28]. New developments in computational PPI prediction have closely mirrored advances in ML and deep learning (DL), especially those of natural language processing.

Beyond furthering our understanding of biological processes at the proteome scale, the ability to accurately predict whether two specific proteins are likely to engage in a physical interaction has significant implications in drug discovery. Indeed, in addition to streamlining the target identification process, this ability promises to significantly accelerate the drug design process itself, notably through the engineering of artificial peptide-protein interactions (PepPIs) [29], [30], [31].

In this review, we turn our attention to sequence-based PPI prediction. First, we make the case for sequence-based approaches and explain why they represent a competitive alternative to structure-based approaches. Second, we review the machine learning methodology used to train and evaluate PPI predictors. Next, we discuss how the lack of training data for non-model organisms is managed and the issue of class imbalance. We then provide the reader with a survey of recent machine learning-based and similarity-based PPI predictors. Subsequently, we describe how sequence-based PPI prediction is reshaping the drug discovery landscape, especially peptide binder and antibody development. Finally, we briefly discuss challenges and future trends within this blooming field of research.

# The case for sequence-based PPI predictors; advantages over structure-based prediction

Modern PPI predictors largely fall into one of three paradigms, depending on the nature of the information they use as inputs: *sequence-based*, *structure-based*, and *hybrid* prediction. Sequence-based predictors utilize the amino acid sequences of the proteins in a pair to make predictions. In contrast, structure-based methods make use of the coordinates of atoms in 3-dimensional space. Hybrid predictors are informed by both sequence and structural information. There is an unresolved dispute among experts surrounding which paradigm shows the most promise.

While structure-based and hybrid methods have done well and may appear more "powerful" at first, since they make use of rich, highly granular information, they are not without their limitations. First, in order to make accurate predictions, these methods require high-quality structures. At the time of writing, the RCSB Protein Data Bank (PDB) [32] contains high-resolution (≤2Å) structures for slightly over 28,200 structures involving 3,772 distinct human proteins, of which only about ~40% are not significantly truncated (>80% of the full-length protein). The growth in available high-quality structures cannot keep up with the growth of experimentally validated PPI databases (Figure 1).

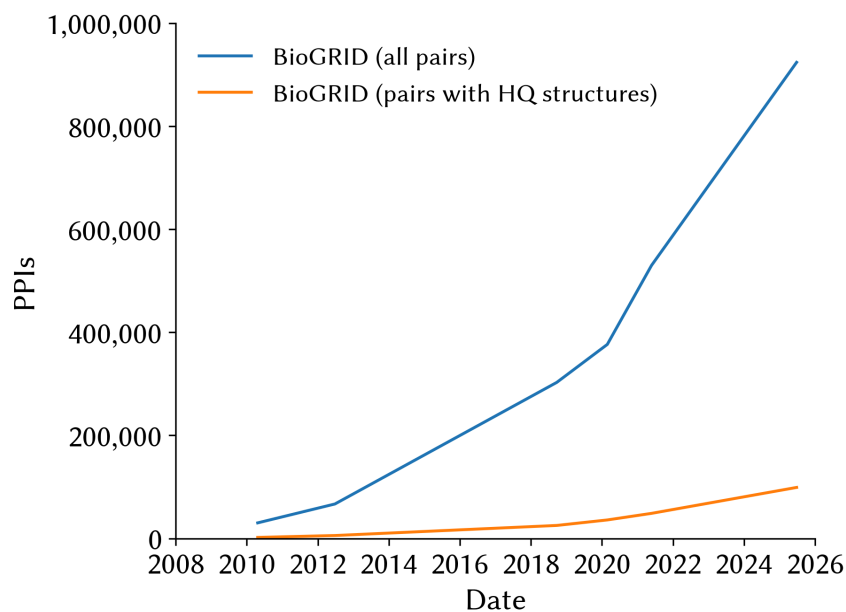

**Figure 1 | Gap between the total number of PPIs and PPIs for which high-quality structures are available**
Only a fraction of the total number of experimentally validated *physical* PPIs in the BioGRID PPI database [33] involving at least one human protein (blue) have high-quality structures (<2Å resolution) for both interactors deposited in the RCSB Protein Data Bank [32] (orange). This fraction of PPIs with high quality structures has been decreasing over time. Furthermore, the fraction of *useful* structures may be lower, since many of these structures were resolved for heavily truncated protein constructs. (Data retrieved and compiled using the RCSB PDB API.)

While models like AlphaFold2/3 [34], [35], ESMFold [36], Chai [37], and Boltz-1/2 [38], [39] have produced comparatively impressive structure predictions, the quality of the predictions vary at the proteome scale. As a result, they are not expected to replace experimental methods such as X-ray crystallography and are said to be most valuable for guiding hypothesis and accelerating early discovery [40]. Furthermore, these tools attempt to model intrinsically disordered regions within proteins which lack a clearly defined structure with limited success [41], [42], [43], [44], [45]. This is significant, given that it is estimated that intrinsically disordered regions represent 30-40% [46] of the human proteome. Finally, even the most accurate protein structure prediction models are also limited in their ability to model proteins whose conformation is dynamic, *e.g.* in response to a switch between the cofactor-free *apo-* and the cofactor-bound *holo-* states. Structure predictors like AlphaFold2 tend to model the most stable domain orientation in proteins which undergo major conformational changes [47]. Taken together, these shortcomings support the argument that while accurate under certain conditions, structure-based approaches are far from being a panacea when it comes to predicting PPIs.

# How PPI Predictors Work: Machine Learning Methods and Evaluation Metrics

## Paradigms

One of the most widely accepted definitions for "machine learning" is that of Tom Mitchell [48]:

> *A computer program is said to learn from experience E with respect to some class of tasks T and performance measure P, if its performance at tasks in T, as measured by P, improves with experience E.*

This definition applies to PPI prediction, as ML-based predictors become more accurate (*P*) at distinguishing between interacting and non-interacting protein pairs (*T*) the more PPIs they are "shown" (*E*).

The PPI prediction challenge is a *binary classification* problem where protein pairs must be assigned to one of two *classes*: *interacting* ("positive") or *non-interacting* ("negative"). Related challenges, such as predicting binding affinities [49], [50], [51], [52] or identifying interaction interfaces [53], [54], [55], [56], have also been explored for two decades.

*Supervised learning* remains the dominant paradigm in PPI prediction, though breakthroughs in deep learning have led to the emergence of a new paradigm with which it is often combined: *self-supervised learning*. Supervised learning is a ML paradigm wherein models are trained using *labeled* data, *i.e.* data for which the target variable to predict is known. Its objective is to discover patterns in data which correlate with the target variable. In the PPI prediction context, because the classes of all protein pairs in the training set are known, the parameters of the model can be tuned so that it identifies and processes the correct patterns to make the best possible predictions.

Self-supervised learning is a relatively new paradigm that arose in the field of natural language processing in response to the availability of colossal quantities of *unlabeled* data (*e.g.* Wikipedia, internet forums, scientific articles and books, corpora of digitized books, *etc.*). This paradigm is typically not used to make predictions directly, but rather to uncover effective ways to distill complex data in a compact and information-rich representations, typically as long vectors of real numbers referred to as *embeddings*. These embeddings are then used as features for various related prediction tasks. Self-supervised learning is now ubiquitous in protein-related ML applications and the embeddings generated in a self-supervised settings are commonly used in conjunction with traditional supervised ML models.

## Methodology

The development of most machine learning (ML)-based predictors, regardless of the application, follows a standard methodology (Figure 2), which we expand on in this section.

### Data curation

Consistent with the "garbage in, garbage out" adage, the creation of high-quality training and test sets is a necessary step towards the creation of a reliable PPI predictor. PPIs are typically retrieved from carefully curated databases which catalog experimentally validated and probable PPIs which are made publicly available for use by biologists, biochemists and bioinformatic practitioners. These databases list physical (direct) and genetic (indirect) interactions to facilitate functional characterization of proteins and drug target identification, among others. The most widely used and recently updated databases are listed in Table 1.

These databases provide the model with a source of interacting pairs (positives) but do not tabulate non-interacting pairs which are required to train binary classifiers. For this reason, a set of non-interacting pairs must be carefully assembled. While this may seem simple at first, it is difficult to prove that two proteins *never* interact. It is possible that a protein pair not currently known to interact may eventually be shown to interact. As a result, it is typical to create negative pairs using one of the following strategies:

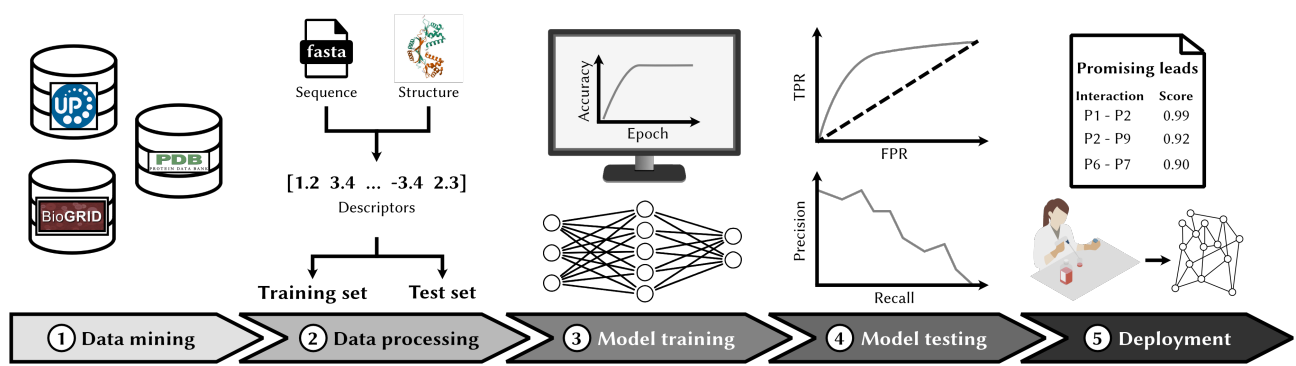

**Figure 2 | Machine learning workflow for PPI prediction**
A standard methodology is typically employed to train ML-based PPI predictors. First, known interactions and their protein sequences are retrieved from databases and curated. Pairs of proteins assumed to be noninteracting ("negatives" pairs) are added to the dataset. Second, numerical descriptors of protein sequences and/or structures are extracted, and the dataset is split into a training and a validation set. Third, a model is trained by minimizing misclassifications over the training protein pairs. Fourth, the model predicts PPIs in a test set composed of new, unseen protein pairs, allowing for an accurate estimate of performance. Finally, the model is deployed to identify new interactions for *in vitro* validation and predict the interactome.

1. Assume random pairs of proteins to not interact [26];
2. Shuffle the amino acids in pairs (*e.g.* in triplets) of interacting proteins, to retain the original amino acid composition [57];
3. Assume pairs of proteins located in different cellular components (*e.g.* cytosol and the nucleus) to not interact [57] (argued to lead to overoptimistic performance estimates due to functional bias [58]).

Regardless of the approach taken, there is a risk of mislabeling a protein pair as negative when they actually *would* interact, though this risk is assumed to be negligible in practice.

Another resource that has been used [59] to gather negative pairs is Negatome [60], [61], a database that lists protein pairs deemed to be unlikely to interact physically. The database was populated using text mining against PubMed-indexed articles and structural information.

In contrast with protein structures, sequences for all known proteins are readily available in databases such as Swiss-Prot/UniProt [62], so the sequences for all positive and negative protein pairs in a dataset can easily be retrieved.

**Table 1 | Recently updated protein-protein interaction (PPI) and peptide-protein interaction (PepPI) databases incorporating experimental data involving human proteins**

| Database | URL | Human interactions |
| --- | --- | --- |
| BioGRID [33] | https://thebiogrid.org | 1,890,522 |
| STRING [63] | https://string-db.org | 2,219,787 |
| IntAct [64] | https://www.ebi.ac.uk/intact | 1,702,367 |
| MINT [65] | https://mint.bio.uniroma2.it | 139,901 |
| Propedia [66] | http://bioinfo.dcc.ufmg.br/propedia | 19,813 |

**Feature engineering and data splitting**

Until the mainstream adoption of deep learning, sequence-based PPI predictors relied on human-engineered (or interpretable) vectors of real numbers as inputs to train supervised models ([Figure 3](Figure 3)A). These vectors, whose components are referred to as *features* or *descriptors*, vary in length and are numerical representations of the properties of proteins in the pair or the pair as a whole. Given that supervised models extract patterns from these descriptors and combinations thereof, the careful design of features that correlate with the target variable is primordial.

At the time of writing, however, representations of proteins and protein pairs are largely *learned*. Large models trained with self-supervised learning ([Figure 3](Figure 3)B) on large datasets comprising millions of protein sequences now generate effective representations of proteins in absence of any external sources of information about the proteins (*e.g.* physicochemical properties, evolutionary information, *etc.*). We discuss specific feature extraction strategies later, in our survey of ML-based PPI predictors.

Regardless of what and how features are extracted from protein pairs to enable classification, the dataset consisting of positive and negative protein pairs is invariably split into a training set and a test set. This can be done in a stratified fashion or not, *i.e.* the positive-to-negative ratio may or may not be the same in the training and the test sets.

The training set is used to tune the model, *i.e.* to find the parameters which allow it to make the best possible predictions on those training pairs. The test set, on the other hand, is set aside early and only used to evaluate the model's predictive accuracy on new, previously unseen data. The accuracy of the model's predictions on a carefully prepared test set provides a measure of how well the model is expected to perform in a "real life" setting.

It is standard practice to correct for high sequence redundancy, as the presence of high similarity sequences introduces bias [67], [68]. A typical way to address the issue of redundancy is to cluster the pairs based on the identity of the interacting proteins [59], [69], [70], [71] with tools such as CD-HIT [68] or MMSeqs2 [72]. A threshold of 40% identity (which allows sequences in the dataset to have ≤40% identity) appears to be commonplace, in practice [26], [59], [70], [73], [74], [75].

Correction for redundancy is not only important for reducing bias during the training of ML-based predictors, but also to obtain accurate estimates of performance. Presence of protein sequences in the test set which share high identity with other sequences in the training set is likely to lead to overly optimistic performance estimates which do not generalize at the proteome scale. This was recognized by Park and Marcotte as early as in 2012 [76]. They argue for the need to distinguish between three classes of protein pairs in the training set: pairs where both proteins are found in at least one interaction in the training set (C1; easiest), pairs where one of the two proteins belongs to an interacting pair in the training set (C2; moderate), and pairs where both proteins do not appear in the training data (C3; challenging). They argue that success in classifying pairs of Class 1 is unlikely to generalize at the proteome scale. So-called "hub proteins" – promiscuous proteins – are especially susceptible to bias performance estimates and require attention when preparing a test set [77].

**Model training**

A variety of standard machine learning models are fit, in isolation or as ensembles, to the training data. Commonly encountered models include multilayer perceptrons (MLPs), support vector machines (SVMs), random forests (RFs), extra trees (ETs), and convolutional neural networks (CNNs).

MLPs are simple neural networks that apply compositions of non-linear operations (*e.g.* sigmoid function, rectified linear unit, *etc.*) to linear combinations of the input features to generate a single real number corresponding to the probability of an interaction. The parameters of the model which are learned during training are the coefficients of the linear combinations. SVMs, on the other hand, project the training pairs into a high-dimensional space and attempt to fit the hyperplane that best separates – *i.e.* with the largest margin – interacting pairs and non-interacting pairs. RFs are ensembles of decision trees where individual trees operate on separate random subsets of the feature space and are trained to find the split points that best separate interacting from non-interacting pairs. CNNs treat protein pairs as an image (a 2- or 3-dimensional tensor) and apply a series of convolution and pooling operations to the images in a way that mimics the way in which the processing of visual information was believed to happen in the primary visual cortex. These algorithms are described in great depth in most introductory machine learning textbooks [78], [79], [80].

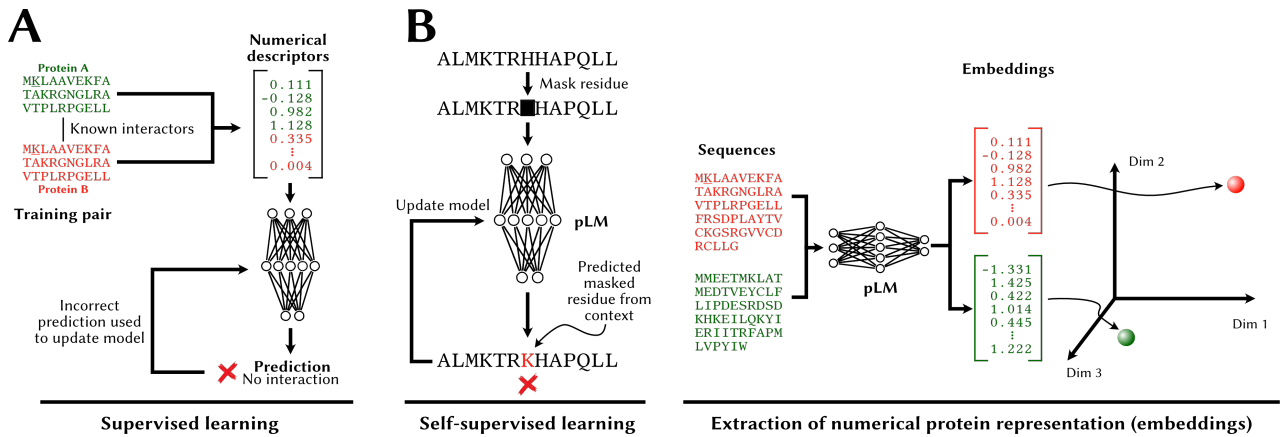

**Figure 3 | Supervised learning and self-supervised learning in PPI prediction**
**(A)** In the supervised learning setting, a model is trained to make predictions on numerical representations (vectors of descriptors/features) representing protein pairs. These descriptors can be human-engineered (traditional ML) or learned from data by deep learning models. The model's parameters are adjusted to maximize the number of correct predictions over a training set. **(B)** In self-supervised learning, a protein language model (pLM) learns to produce representations (embeddings) for proteins by learning to predict a masked residue from surrounding amino acids (masked language modeling).

The specific algorithmic details pertaining to how each of these models are trained are beyond the scope of this review. Suffice to say, the training procedure is, invariably, an optimization routine that minimizes an objective function (also referred to as a *loss function*), typically some form of misclassification error aggregate over the training set.

**Model evaluation**

Several standard metrics are used to assess the quality of a PPI predictor (Figure 4). Several of the metrics, which vary between 0 and 1, are formulated as ratios of true positives (TP), true negatives (TN), false positives (TP), and false negatives (FN) from predictions made on the test set of protein pairs that weren't used to train the model.

Recall (also known as *sensitivity*) quantifies a predictor's ability to detect interactions:

$$\text{Recall (Re)} = \frac{\text{TP}}{\text{TP} + \text{FN}}$$

In contrast, specificity provides a measure of how well the predictor can detect non-interacting protein pairs:

$$\text{Specificity (Sp)} = \frac{\text{TN}}{\text{TN} + \text{FP}}$$

Precision, arguably the performance metrics which matters the most when the predictor is deployed to validate novel PPIs *in vitro*, provides the expected fraction of predicted interacting pairs which would be confirmed to interact upon testing:

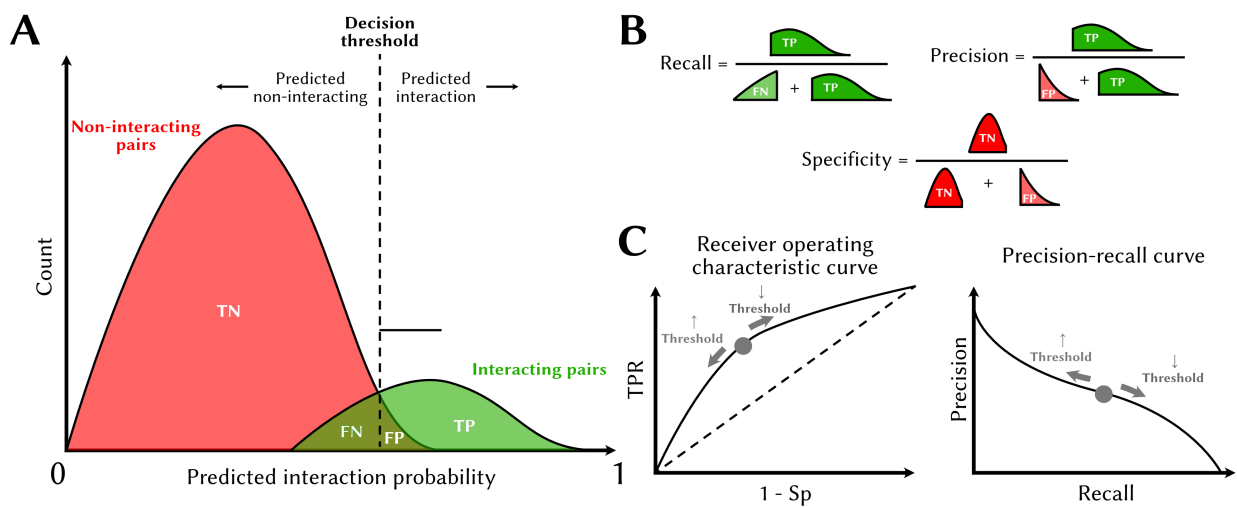

**Figure 4 | Assessment of a PPI predictor performance**
**(A)** Schematic representation of the score distributions for non-interacting (red) and interacting (green) pairs. Protein pairs with scores/probabilities above the arbitrarily set decision threshold are classified as positive. **(B)** Schematic representation of how the score distributions and predictions are used to compute important performance metrics. **(C)** The receiver operating characteristic curve (left) illustrates the tradeoff between the true positive rate (TPR) and specificity (Sp) over the range of decision thresholds. The ROC curve of the perfect PPI predictor passes through the upper-left corner of the plot. The precision-recall curve (right) illustrates the tradeoff between precision and recall over the range of decision thresholds. The PR curve of the perfect predictor passes through the upper-right corner of the plot.

$$\text{Precision (Pr)} = \frac{\text{TP}}{\text{TP} + \text{FP}}$$

The $F_1$-score is sometimes found to be convenient as it captures both the recall and precision of a predictor in a single number by means of a harmonic mean:

$$F_1\text{-score} = 2\frac{\text{Pr} \times \text{Re}}{\text{Pr} + \text{Re}}$$

Accuracy is rarely used, for reasons that are discussed later, but it quantifies the fraction of expected correct predictions:

$$\text{Accuracy} = \frac{\text{TP} + \text{TN}}{\text{TP} + \text{TN} + \text{FP} + \text{FN}}$$

ML-based predictors output a score – almost always a probability between 0 and 1. The closer to 0 the score is, the more confident the predictor is that the protein pair does not interact, while the opposite is true as the score approaches 1.

The metrics above require a user to define a decision threshold on the score above which protein pairs are predicted to interact. This threshold is arbitrary, but is selected to achieve the desired balance between recall and precision, or less frequently, recall and specificity. Alas, improved precision incurs a cost in recall and vice-versa. Two metrics that are frequently used to report a model's performance over the range of possible threshold values are (1) the area under the receiver operating characteristic curve (AUROC) and (2) the area under the precision-recall curve (AUPRC), also sometimes called "average precision". The AUROC and AUPRC are useful to compare the performance of different predictors, as are metrics such as the precision at a fixed recall value (*e.g.* the precision at 50% recall; Pr@50Re).

# 4. Generalizing Beyond Model Systems: Challenges and Solutions in Cross-Species PPI Prediction

The literature distinguishes between three main prediction schemes: *intra*-, *inter*-, and *cross*-species PPI prediction (Figure 5). The distinction is important as species are thought differ from one another with respect to their interaction patterns [27], [81].

Intra-species prediction is the most common prediction scheme wherein one seeks to predict a full or partial interaction network within a single organism, using known interaction within that organism.

By contrast, in the inter-species prediction scheme, predictions involving proteins from different organisms are made. This scheme has received significant attention because of the COVID-19 pandemic where the interaction of SARS-CoV-2 proteins and human proteins, notably Spike and ACE2, were determined to be key to the infection and proliferation of the virus. Human-virus protein interaction prediction with sequence-based PPI predictors has been the topic of multiple studies since [82], [83], [84], [85]. Inter-species PPI prediction has also been used to predict interactions between soybean (*Glycine max*) and the soybean cyst nematode (*Heterodera glycines*) [86], a parasite that contaminates crops and leads to millions of dollars in yield losses as well as the human-HIV interactions [87], to name a few.

Unfortunately, the scarcity of training data (*i.e.* known, validated PPIs) for the organism of interest is a frequently encountered problem. Cross-species PPI prediction, which differs from inter-species prediction, is the most frequently used strategy to mitigate this issue [27], [81], [88], [89]. In cross-species prediction, a well-studied organism closely related to the target organism is taken as a "proxy" [81] for the target organism and PPIs from the proxy organism are used to inform the predictor. For instance, at the beginning of the COVID-19 pandemic, few interactions between SARS-CoV-2 proteins and human proteins had been experimentally demonstrated. Dick *et al.* used 689 PPIs involving proteins from closely related viral families (*Flaviviridae*, *Togaviridae*, *Arteriviridae*, *Coronaviridae*, and *Hepeviridae*) to inform their predictions [81].

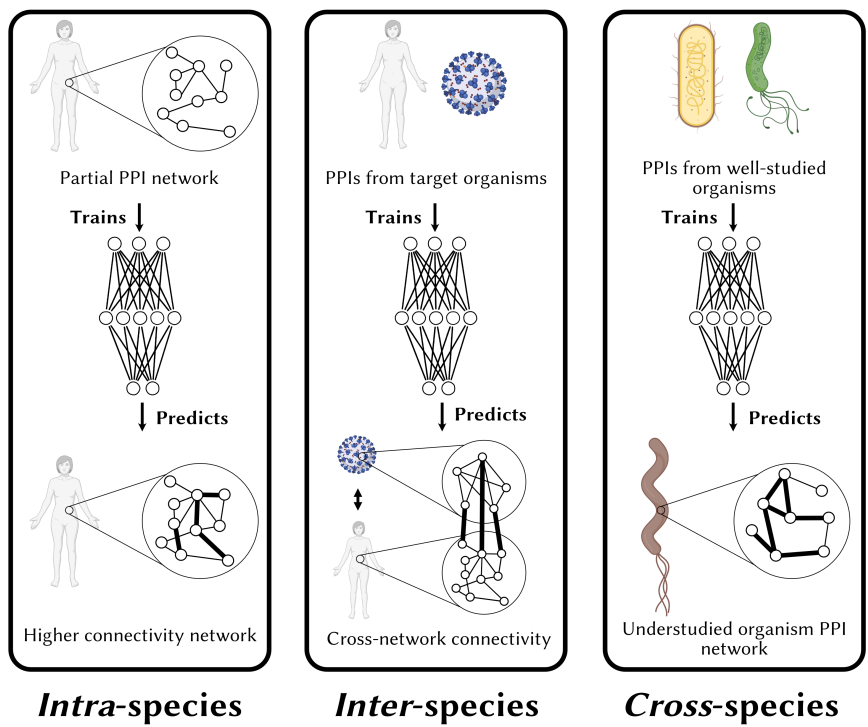

**Figure 5 | Frequently encountered PPI prediction schemes**
In intra-species predictions, PPIs from an organism are used as training data to discover new interactions within the same organism. In the inter-species scheme, interactions from two or more organisms are used to predict interactions between the organisms. Cross-species PPI prediction involves predicting interactions within understudied organisms from PPIs from better studied, evolutionarily-related organisms which act as "proxies".

In a recent publication, Volzhenin *et al.* considered, with impressive minutiae, how their predictor named SENSE-PPI performs under three different prediction schemes [90]. Unsurprisingly, they confirmed that the quality of the predictions in the cross-species schema is inversely correlated with the phylogenetic distance between the proxy organisms and the target organism for which PPIs are to be predicted.

## The class imbalance problem in PPI prediction

As is the case for many binary classification problems in bioinformatics such as miRNA prediction [91], posttranslational modification (PTM) prediction [92], and antimicrobial peptide prediction [93], PPI prediction is plagued with the *class imbalance problem*. This occurs when instances of one class, typically the "negative" class, vastly outnumber instances of the rare class, the "positives".

It is reasonable for a researcher to wonder why this matters, especially since many published predictors fail to account for this imbalance. The answer is that testing a model on a balanced test set is a recipe for unrealistic and overly optimistic performance estimates, because a balanced test set is not a representative sample of the actual, imbalanced distribution that is to be expected upon deployment. Mitigating the effects of class imbalance involves two things: the use of an imbalanced test set to evaluate the predictor and the use of appropriate performance metrics.

In general, though not always, the higher the expected imbalance is, the more challenging the classification task is [94], so it is essential to test the predictor on a test set which is as challenging as the set of all protein pairs. Testing a predictor on a balanced dataset would be akin to administering a high school-level exam to a graduate student: the results would not provide useful information about the graduate student's *expected* performance upon graduation from post-secondary school. The use of appropriate performance metrics is another way to mitigate the effects of class imbalance. Several metrics widely adopted for benchmarking are unsuitable or largely uninformative for the evaluation of PPI predictors: accuracy and the AUROC.

Accuracy is not an adequate metric as it disproportionately favors correct classification of the majority class. For example, assuming that 1% of all protein pairs actually interact, a predictor which consistently predicts pairs as non-interacting would achieve an accuracy of 99%. Clearly, this predictor is of no practical use, despite its very high accuracy.

The ROC curve and the area under it, in contrast with the PR curve, is insensitive to class imbalance [74], [95]. As such, it does not correlate with the difficulty of the classification problem at different imbalance ratios. We illustrate these phenomena in with simulated datasets in Figure 6.

In general, the metrics that are the most relevant in presence of high class imbalance are (1) recall, (2) precision and (3) their derivatives, AUPRC and $F_1$-score; we are mainly interested in detecting positive with high sensitivity and making correct positive predictions [96]. Rarely are we ever interested in correct classification of negatives in the context of PPI prediction. Some predictors developed in imbalanced settings report the prevalence-corrected precision, which can be used to estimate precision under different imbalance ratios [91], [97]:

$$\text{PCPr} = \frac{\text{Re}}{\text{Re} + r(1 - \text{Sp})}$$

where Re is the recall, Sp is the specificity, and $r$, the prevalent-to-rare instance ratio. As an example, $r = 100$ in a scenario where a 1:100 ratio is expected.

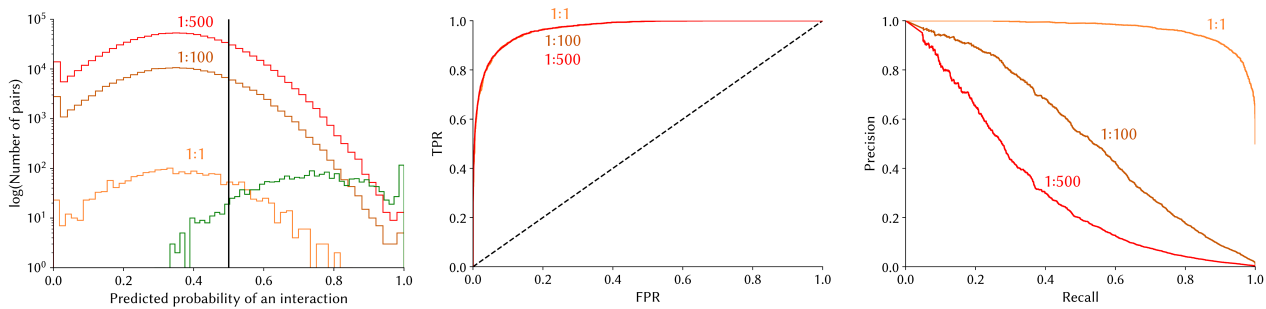

**Figure 6 | Sensitivity of the ROC and PR curves to class imbalance**
In this simulated scenario, where the scores of interacting and non-interacting protein pairs are assumed to be normally distributed, we compare the ROC and PR curves obtained for different class imbalance ratios (1:1, 1:100, and 1:500).

Though the ratio between interacting and non-interacting protein pairs is difficult to approximate, it is widely accepted that the majority of protein pairs do not interact. In a 2009 study [98], Venkatesan *et al.* predicted based on the results of a Yeast Two-Hybrid assay variant they developed that the human interactome contains somewhere between 74,000 and 200,000 interactions (*i.e.* a~1:1,200-3,400 imbalance ratio). Stumpf *et al.* [99] estimated the size of the human interactome at 650,000 interactions (*i.e.* a ~1:400 imbalance ratio) with a graph-based approach. PPI predictors are typically tested under much lower imbalance regimes such as 1:10 [27], [70], [74] or 1:100 [26]. This suggests that the performance of the predictors may be consistently overestimated.

# Old but Gold: Sequence-Based Protein-Protein and Peptide-Protein Predictors

## Machine learning-based approaches

As stated previously, many PPI predictors employ standard, out-of-the-box machine learning models or combinations thereof. The most noteworthy difference between most predictors is the approach used to generate the protein representations, *i.e.* feature vectors, which is the focal point of interest in this section.

The first widely acknowledged sequence-based predictor was published by Guo *et al.* in 2008 [57]. Their model used 7 physicochemical properties of the side chains as features: hydrophobicity, hydrophilicity, side chain volume, polarity, polarizability, solvent-accessible surface area, and net charge index. To account for the fact that proteins vary in length, these features are aggregated along the length of the protein into a single number using the auto-covariance (AC) aggregator to generate a 7-dimensional protein representation. The authors describe AC as "the average interactions between residues, a certain lag apart throughout the whole sequence". The two vectors for a protein pair are then concatenated to generate a single 14-dimensional vector. The classification model is a SVM trained on protein pairs represented as 14-D vectors. This work was highly influential, especially for its use of the AC to generate a feature vector with a fixed length from protein sequences that vary in length, as well as for the strategy employed to generate non-interacting pairs.

Since then, a wide variety of human-engineered features to describe protein sequences as vectors of real numbers have been proposed. The amino acid composition (AAC) is simply a 20-D vector where each entry corresponds to the percentage of one of the 20 amino acid within the sequence. Chou proposed the pseudoamino acid composition (PseAAC) [100], a variation on AAC which aims to preserve some of the sequential information in the protein sequence by adding to the AAC vector factors that account for the correlation of physicochemical properties of sets of amino acids spaced 1, 2, 3, ..., $\lambda$ amino acids apart. PseAAC generates a $(20 + \lambda)$-D vector.

Another frequently encountered feature vector encountered in PPI prediction is that output by the conjoint triad (CT) method [101]. In the CT method, each of the 20 possible amino acid are assigned to one of 7 groups whose members share similar physicochemical properties (side chain volume and dipole). The counts of each possible group triplet in the protein sequence make up the entries of the $7 \times 7 \times 7 = 343$-D feature vector which is subsequently min-max normalized.

The composition, transition and distribution (CTD) method [102] consider several properties (7 in the original paper), each of which is split into 3 group of amino acids (*e.g.* the "neutral", "negative", and "positive" groups for the charge property). The "composition" feature corresponds to the frequency of each of those groups in the sequence. The "transition" feature measures the frequency at which an amino acid of one group is followed by an amino acid of another group (a so-called "transition"). Finally, the "distribution" features represent the sequence lengths required to contain the first, 25%, 50%, 75% and 100% of the amino acids of a particular group. The CTD method with 7 properties and 3 groups per property yields a 441-D vector.

Another common numerical description of a protein sequence is the position-specific scoring matrix (PSSM). This $20 \times L$ matrix is generated by running the PSI-BLAST program [103] against a large database of protein sequences to compute the probability of observing all 20 amino acids at a given position in an alignment of high-scoring BLAST hits obtained from the input sequence for each of its $L$ residues. The PSSM captures important conservation patterns between the protein of interest and other proteins.

Romero-Molina *et al.* used ProtDCal [104] to generate tens of thousands of descriptors, of which a small number were selected to train a SVM model. ProtDCal applies a variety of grouping schemes, weights, and aggregation operations on the physicochemical properties of the residues in the input sequence to generate features that represent the protein globally and locally.

The Word2Vec model [105] was also used by some groups [83], [106] to generate protein representations which capture context. The Word2Vec model, initially developed for natural language processing, can be used to generate protein embeddings if it is trained on large numbers of protein sequences. Two strategies to train Word2Vec are used: skip-gram, where the model learns representation by being trained to predict context from a word, and continuous bag of words, where the model learns representations by being trained to predict a missing word from context.

The models trained on combinations of these representations to predict PPIs include CNNs [69], [107], [108], [109], [110], [111], SVMs [57], [59], [112], RFs [73], [82], [113], MLPs [111], [114], and ensembles of such models [73], [84], [113].

**Protein language model-based approaches**

The publication of the transformer neural network architecture by Google in the 2017 paper, *Attention Is All You Need* [115], was nothing short of transformative for the field of natural language processing (NLP). The multibillion-parameter large language models (LLMs) which we are now familiar with at the time of this publication, such as OpenAI's ChatGPT [116] and Google's Gemini [117], invariably build on top of this architecture.

What makes transformer-based models so powerful for modeling language is the use of the *self-attention mechanism* which allows models to attend more to relevant and/or related elements when processing a sequence. For instance, let us consider the natural language sentence: *"Peptide inhibitors are a promising therapeutic modality"*. When embedding the word "Peptide" into the representation of the sentence, the model would consider the strength of its relationship with the words "inhibitors" and "promising" as stronger than with the words "are" or "a". This mechanism allows transformers to better capture the relationship between words in a sequence than other deep learning models such as long short-term memory networks (LSTMs) could previously.

The idea of treating protein sequences as a language with its own vocabulary, semantics, and grammar is a relatively recent idea, even if statistical models such as Hidden Markov Models (HMM) have been used for homology modeling and sequence retrieval in databases for decades [118], [119], [120]. Before the emergence of the transformer, LSTMs were used to model the protein language [121], but they have since fell out of favor and were replaced with transformer-based models, except for applications where little training data is available. LLMs architectures trained to model the language of protein sequences are referred to as protein language models (pLMs) and have become a mainstay in PPI prediction and protein bioinformatics more broadly.

The main purpose of pLMs is to generate rich, high-dimensional embeddings of protein sequences which indirectly capture a sequence's physicochemical, evolutionary, functional and structural information. These models typically learn effective protein representations through the masked language modeling task where one or multiple residues within tens of millions of protein sequences are masked and the model is tasked to correctly predict the missing residues based on the context (*i.e.* the other known residues in the sequence).

A number of "foundational" pLMs trained at great expense on very large sets of pre-clustered, non-redundant protein sequences such as UniRef [122] or the "Big Fantastic Database" [123], are publicly available (Table 2). These models can be used to generate representations we can be used as-is for downstream classification challenges with traditional models (*e.g.* SVMs, MLs, CNNs, RFs, *etc.*) or fine-tuned on additional data to generate protein sequences having desired properties, antimicrobial peptides [124], [125] for example.

Over the last couple of years, pLM-generated representations have been increasingly used to predict PPIs. For instance, Hu *et al.* trained their CNN-based model, KSGPP [110], on ESM-2 protein embeddings combined with a graph-based representation of the STRING PPI network. Another model, TuNA [126], uses also uses ESM-2 embeddings which are further processed in another transformer network and classified with a Gaussian process classifier. The xCAPT5 model by Dang *et al.* [111], uses the embeddings produced by the ProtT5-XL model trained on UniRef50 as inputs to their siamese CNN model.

**Table 2 | Foundational protein language models used to predict PPIs**

| Model | Embedding dimension | Parameters (approx.) | Training strategy | Training data |
|---|---|---|---|---|
| ProtT5 [127] | 1,024 | 3B | 1-gram random masking with demasking | BFD (pre-training; ~2.1B sequences) and UniRef50 (fine-tuning; ~45M sequences) |
| Ankh [128] | 1,536 | 1B | 1-gram random masking with full sequence reconstruction | UniRef50 (~45M sequences) |
| ESM-2 [36] | 1,280 | 650M | 1-gram random masking with demasking | UniRef50+90 (~65M sequences) |

## Similarity-based approaches

The prediction of the comprehensive human interactome had long been intractable before the development of a "massively-parallel" implementation of the PIPE algorithm (MP-PIPE) [129]. Thanks to its ability to predict whether a protein pair will interact in a fraction of a second, MP-PIPE was deployed on the 250M pairs of human proteins and provided the first map of the human interactome in 2011. This feat, which involved a significant amount of compute (~3 months on a cluster with 800 compute cores or ~170,000 CPU-hours), revealed 130,470 high-confidence novel interactions (when setting the threshold at a 0.05% false positive rate). The Scoring Protein INTeractions (SPRINT) algorithm [26] was demonstrated to have the ability to predict the human interactome in a matter of hours with a consumer-grade workstation in 2017.

More recently, deep learning models were used to make predictions for large numbers (*i.e.* tens of millions) of pairs, but these methods first applied filters to reduce the number of predictions to make (*e.g.* only scoring pairs with a common subcellular localization [130]). Because of MP-PIPE and SPRINT's unique ability to predict entire interactomes, it is worth discussing these two PPI prediction algorithms which belong to a family of methods referred to as "similarity-based methods".

Similarity-based methods rely on gap-free alignment scores obtained with substitution matrices like PAM120 [131] as a measure of similarity between the windows of contiguous amino acids within proteins.

The fundamental idea underpinning these methods is that protein interactions are mediated by short windows of contiguous amino acids. Clearly, this working assumption is incorrect as residues located at the interface which mediate the interaction may be far apart in the protein's sequence but proximal in 3-D space as a result of protein folding. Nonetheless, similarity-based methods have been surprisingly useful, nonetheless. To quantify the evidence in favor or a putative interaction, these methods posit that a pair of known interacting partners, ($I_1$,$I_2$), provides evidence for an interaction between a pair of query proteins, ($Q_1$, $Q_2$) if $Q_1$ contains a short window similar to a window in $I_1$ and $Q_2$ contains a short window which is similar to a window in $I_2$.

The various implementations of PIPE [27], [129], [25] and SPRINT merely quantify the evidence by counting the number of similarity occurrences with proteins in an interaction list. While PIPE outputs a score manipulated to lie in the range [0, 1], SPRINT was not and outputs a score in the interval [0, ∞), and because they essentially produce counts, the scores output by PIPE and SPRINT should not be interpreted as interaction probabilities. Instead, they should be thought of as the strength of evidence in support of an interaction. Since similarity-based methods produce counts of similarity with interacting proteins, the concept of non-interacting pair is foreign to these algorithms and they therefore do not require a set of noninteracting protein pairs to make predictions.

PIPE and SPRINT are very similar, but differ in a few key aspects, some of which pertain to computational efficiency while others pertain to the scoring itself. In contrast with PIPE, SPRINT does not perform an exhaustive search for regions of similarity between proteins. Instead, it uses spaced seeds to coarsely look for potential regions of similarity to be assessed for similarity with more scrutiny. This heuristic allows SPRINT to achieve impressive speed gains over PIPE, as identifying the regions of similarity between proteins is the most computationally intensive step of both algorithms. Second, PIPE uses a simple threshold and does not account for the level of similarity between windows in query proteins and interacting proteins while SPRINT weighs the counts with the alignment scores.

Because these methods have no awareness of the concept of a non-interacting protein pair and are not *intrinsically* classifiers, one must set a threshold on the score below which two proteins are assumed to not interact. One way to achieve this is to determine the score threshold which achieves a certain sensitivity and/or specificity in a cross-validation experiment [25], [129]. That said, Dick *et al.* made the argument that a global threshold applied to all protein pairs may not be appropriate and suggested a meta-classifier called Reciprocal Perspective (RP) [97]. RP considers the interaction scores of the query proteins with one another among all scores from the perspective of both query proteins and outputs a revised interaction scores on which a global threshold can be set. The use of RP significantly leads to more sensitive and precise predictions than is possible with the use of a global threshold on the original interaction scores.

In spite of their relative simplicity and the incorrect assumption the interactions are mediated by contiguous amino acids, PIPE and SPRINT remain competitive to this day [28] and newer model are still frequently compared against them [28], [75], [77], [89], [132].

**Table 3 | Summary of sequence-based PPI predictors published in the last decade**

| Predictor | Year | Model | Feature set | Main dataset (human) | Negatives | Imbalance |
|---|---|---|---|---|---|---|
| DeNovo[a] [112] | 2016 | support vector machine (SVM) | CT | 5,445 PPIs between human and viral proteins from 173 viruses (from VirusMentha [133]) | Random pairings of proteins not known to interact, excluding pairings where the viral protein is >0.2 similar to another viral protein which interacts with the human protein | 1:1 |
| SPRINT [26] | 2017 | Similarity comparison with PAM120/BLOSUM64 matrix | Not applicable (similarity-based) | 215,029 PPIs from BioGRID [33] | Not needed | 1:100 |
| DPPI [107] | 2018 | Siamese convolutional neural networks (CNNs) | PSSM generated with PSI-BLAST [103] | PPIs from DIP [134]; and 289,180 PPIs from HINT [135] | Random pairings of proteins not known to interact | 1:10 |
| PIPR [69] | 2019 | Siamese CNNs | Concatenation of 1) co-occurrence similarity of the amino acids with skip-gram model and 2) 7-D one-hot vector to denote the amino acid cluster (dipoles and side chain volume clustering) to with the amino acid belongs | 148,051 PPIs from the STRING database [63] | Random sampling of pairs whose proteins must have different subcellular localizations | 1:1 |
| DEEPFE-PPI [106] | 2019 | multilayer perceptron (MLP) | Word2Vec embedding of amino acid sequences with a skip-gram model (context window of length 4) trained on 558,590 sequences from the Swiss-Prot database | 3,899 PPIs from the HPRD database [136] | Unclear | 1:1 |
| PPI-Detect [59] | 2019 | SVM | 19 features selected from 13,248 of protein descriptors generated with the ProtDCal software [104] | 1,922 PPIs from the 3did database [137] and iPfam [138] | Negatome 2.0 [61] | ~1:2 |
| PIPE4 [27] | 2020 | Similarity comparison with PAM120/BLOSUM64 matrix | Not applicable (similarity-based) | 66,084 PPIs from BioGRID (filtered with Positome [139]) | Not needed | 1:10 |
| StackPPI [73] | 2020 | Ensemble of random forests (RFs) and extra trees (ETs) feeding into and a logistic regression meta-classifier | PseAAC, AC (3 properties), PSSM, CTD | 5,594 PPIs from the DIP database; filtered to retain a maximum of 40% identity between interactors | Random pairings of proteins not known to interact | 1:1 |
| GTB-PPI [71] | 2020 | Gradient-boosted trees | 331 features selected among PseAAC, pseudo-PSSM, reduced sequence and index-vectors, and AC | 1,412 human PPIs used for testing only (from [140]) | Unclear | 1:1 |
| InterSPPI[a] [82] | 2020 | RF | CT, CTD, AC (Guo's 7 physicochemical properties [57]) | 22,653 PPIs from the HPIDB database (v3.0) [141] | Same as DeNovo (see above) | 1:10 |
| LSTM-PHV[a] [83] | 2021 | long short-term memory network (LSTM) | Word2Vec (Continuous Bag-of-Words Model) trained on the Swiss-Prot database with a context of 4 amino acids to produce 128-dimensional embedding vectors which are concatenated to produce the embedding matrixes of proteins | 22,383 human-virus PPIs from the HPIDB database (v3.0) [141], 7,373 human-SARS-CoV-2 PPIs from BioGRID | Variation on DeNovo's approach (see above) | 1:10 |

| Predictor | Year | Model | Feature set | Main dataset (human) | Negatives | Imbalance |
|---|---|---|---|---|---|---|
| TransPPI[a] [84] | 2021 | Siamese CNNs and MLP | PSSM generated with PSI-BLAST [103] | 31,381 viral PPIs from HPIDB, VirHost-Net [142], VirusMentha, PHISTO [143] and PDB [32]; 568 human-SARS-CoV-2 interactions from MS experiments [144], [145] | Variation on DeNovo's approach (see above) | 1:10 |
| D-SCRIPT [70] | 2021 | Complex multimodule network with convolution components | Embeddings produced with a Bi-LSTM developed by Bepler and Berger [146] | 47,932 PPIs from the STRING database [63] | Random pairings of proteins not known to interact | 1:10 |
| Deep-Trio [108] | 2022 | CNN | One-hot encoding | 31,164 PPIs from BioGRID | Shuffling one sequence of an interacting pair with 2-let counts (excluding the first residue of the protein), suggested in [57] | 1:1 |
| SDNN-PPI [114] | 2022 | MLP with self-attention layers | AAC, CT, AC (with Guo's 7 properties) | 3,899 PPIs from HPRD | Random sampling of pairs whose proteins must have different subcellular localizations | 1:1 |
| Topsy-Turvy [88] | 2022 | Combination of D-SCRIPT and a graph spectral theoretic module | Embeddings produced with a Bi-LSTM developed by Bepler and Berger [146] (for D-SCRIPT component) and a partial PPI network for the graph spectral theoretic module | 47,932 PPIs from STRING | Random pairings of proteins not known to interact | 1:10 |
| EResCNN [113] | 2023 | A ensemble of ET, RF, LightGBM, CNNs, and XGBoost | PseAAC, multiple mutual information, PSSM, Guo's AC, CT, and "encoding based on grouped weight" | 1,412 human PPIs (for testing only; source unclear) | Random sampling of pairs whose proteins must have different subcellular localizations | 1:1 |
| ProtInteract [109] | 2023 | CNN | Embeddings are generated with a temporal convolutional network autoencoder trained on matrices where each column contains 10 physicochemical properties for the corresponding amino acid in the sequence | 80,000 PPIs from the STRING database | Negatome 2.0 | 1:1 |
| KSGPPI [110] | 2024 | CNNs | ESM-2 embeddings combined with composition of k-spaced amino acid pairs (CKSAAP) [147] and Node2Vec graph encoding of proteins within the STRING PPI network | 8,798 PPIs from the DIP database | Rational graph-based approach | 1:1 |
| TuNA [126] | 2024 | Transformer architecture with a Gaussian process classification layer | ESM-2 embeddings further processed with transformers to generate intra-protein and inter-protein representations | Same dataset as D-SCRIPT | Same as D-SCRIPT | 1:10 |
| xCAPT5 [111] | 2024 | Siamese CNNs, followed by an MLP and XGBoost | Embeddings produced with the ProtT5-XL-UniRef50 pLM | 27,593 PPIs from the Pan dataset [148] | Random sampling of pairs whose proteins must have different subcellular localizations | 1:1 |
| PLM-interact [74] | 2024 | ESM-2 fine-tuned for PPI prediction | Not applicable | Same dataset as D-SCRIPT | Same as D-SCRIPT | 1:10 |

| Predictor | Year | Model | Feature set | Main dataset (human) | Negatives | Imbalance |
| --- | --- | --- | --- | --- | --- | --- |
| SENSE-PPI [90] | 2024 | Siamese GRU Module | Embeddings produced by ESM-2 | 86,000 PPIs from STRING | Random pairings of proteins A and B where: (1) A and B are not known to interact in STRING, (2) no homolog of A is known to interact with a homolog of B at more than 40% sequence identity in STRING, and (3) no homolog, at more than 40% sequence identity, of a known interactor of B is known to interact with a homolog of A. | 1:10 |
| INTREPP-PID [89] | 2024 | Ensemble of 5 averaged weight-decayed LSTMs | Learns its own representations | 24,456 high-confidence (score >0.9) PPIs from STRING carefully curated to follow Park and Marcotte's [76] guidelines | Random pairings of proteins not known to interact | 1:1 |

# Protein-protein interaction prediction for drug development

## Identification of drug targets with PPI network analyses

The prediction and validation of novel PPIs with *in silico* and *in vitro* approaches and the mapping of PPI networks (local and full interactomes) grant us with the ability identify proteins and PPIs to target and, consequently, to modulate biological pathways and treat diseases.

A number of network-based approaches operating on PPI networks have been developed for the specific purpose of identifying drug targets. These approaches represent PPI networks as graphs and apply graph theory metrics (*e.g.* degree, betweenness, distance, eccentricity, modularity, coreness, *etc.*) and algorithms (*e.g.* shortest distance) to identify proteins which can be targeted [149], [150]. The analysis may include constraints on connectivity aiming to minimize potential side-effects resulting from the changes to the network associated with the disruption of a PPI. Such approaches have been leveraged to identify targets in glioblastoma [151], depression [152], and cancer [153].

The details of these approaches are out of scope. More thorough treatment of PPI network-driven target identification can be found in other review articles [149], [154], [155].

## Targeting PPIs with peptide binders

One of the reasons that explains why small molecules have prevailed over peptides up to this point is that peptides tend to have much lower bioavailability, meaning that they reach the target site of action in lower quantities due to stability issues, and, as a result, tend to be much more difficult to successfully deliver, especially through the oral route of delivery [156]. Significant progress in peptide delivery have been achieved. For example, chemical modifications of peptides can be used to enhance their stability while advanced delivery vehicles such as implants, nanoparticles, gels, or emulsions can control the release of peptides into the blood or tissues [157]. Thanks to these advances, peptide therapeutics are now considered an "emergent therapeutic approach" [158].

Peptides have several advantages over small molecules for the disruption of PPIs:

1. **Specificity:** Peptides can be designed such that they bind to a target protein and few off-target proteins. This is a highly desirable property, as off-target interactions can lead to undesired side effects and limit the usefulness of a drug. In general, specificity is considered to be more difficult to achieve with small molecules. For example, side effects such as cytotoxicity is observed with tyrosine kinase inhibitors used to treat cancers as a result of their low specificity [156].

2. **Ease of synthesis:** While certain small molecules predicted to have desirable properties may be difficult or impossible to synthesize, the entire (linear) peptide space is chemically accessible because peptides are simply chains of amino acids which can be linked with well-understood chemistry. Modern peptide synthesis methods such as solid-phase peptide synthesis [159] makes screening of large peptide libraries possible. Peptides can also be modified in a number of ways: cyclization, stapling, lipidation, *etc.* to enhance their pharmacological properties (*e.g.* stability, non-fouling, specificity, *etc.*).

3. **Wider target range:** Thanks to their larger sizes, peptides can be used to target large surfaces that small molecules cannot and inhibit PPIs which typically involve large and shallow surfaces [156].

Because peptides are short proteins, sequence-based PPI predictors can be leveraged directly or indirectly to design binders of proteins involved in a PPI one wishes to disrupt and several groups have capitalized on that fact.

InSiPS [160], [161] is a genetic algorithm developed at Carleton University which evolves peptide binders against a specific target. InSiPS uses the interaction scores output by MP-PIPE, which is informed by validated PPIs, as a measure of peptide quality ("fitness"). More specifically, it attempts to design a peptide which has a high interaction score with the target protein and low interaction scores with off-target proteins. InSiPS defines peptide fitness as

$$f = s_{\text{target}}\bigl[1 - \max\bigl(S_{\text{off-target}}\bigr)\bigr]$$

where $s_{\text{target}} \in [0, 1]$ is the interaction score output by MP-PIPE and $S_{\text{off-target}}$ is the set of scores between the peptide and all off-target proteins. In 2022, InSiPS was used to design peptide binders that interact with SARS-CoV-2's Spike protein with an dissociation constant in the nanomolar range [162]. To our knowledge, InSiPS is the only published peptide binder design algorithm which accounts for and attempts to simultaneously minimize off-target interactions.

The CAMP model [163], a sequence-based PepPI predictor which combines CNNs and self-attention modules for both the peptide and the protein. While it has not been used specifically for the purpose of designing peptide binders, one could apply it in a way similar to InSiPS.

The Chatterjee group at Duke University have proposed several sequence-based generative tools which rely on PPIs and/or PepPIs to design peptide binders of proteins. For instance, Cut&CLIP [164] applied a contrastive learning strategy. First, it embeds peptide fragments extracted from known interacting proteins and the target protein's sequences with ESM-2. Then, they train peptide and protein encoders to re-embed into a common latent space where the embeddings of interacting peptide and protein are in close proximity while the embeddings of non-interacting peptides and proteins are far apart.

The same group published PepMLM [165], an ESM-2-based transformer model that was fine-tuned to design peptide binders by framing the problem as a masked language modeling problem. In other words, PepMLM was tasked with predicting masked residues (*i.e.* the peptide binder's sequence) through the use of context, *i.e.* the target's sequence. PepMLM was trained on 10,000 PepPIs curated from the Propedia database [66]. With their method, they were able to generate a ubiquibody that successfully led to the degradation of βcatenin *in vivo*.

PepPrepCLIP [166] was trained on 11,597 PepPIs extracted from the PDB with a contrastive learning approach similar as that in Cut&CLIP. The main difference lying in how the peptides are generated. In PepPrepCLIP, the peptides are generated by adding gaussian noise to existing peptide binder embeddings in the ESM-2 latent space which are subsequently decoded back into sequence space. Using PepPrepCLIP, the authors were able to generate ubiquitin-bound peptides ("ubiquibodies") that interacted with β-catenin and tagged it for proteolysis.

Taken together, this demonstrates that sequence-based PPI predictors and generative models trained on PPIs/PepPIs are showing great promise for designing peptide-based therapeutics which are currently in high demand.

**Antibody design**

Attempts have also been made to leverage sequence-based models to design antibody-based therapeutics which can disrupt PPIs. These sequence-based models, trained on tens of thousands of antibodies, are almost without exception variants of well-known pLM architectures. These models are especially useful to optimize the sequence of antibodies against a known target protein, *i.e.* to increase the affinity of an existing PPI.

AntiBERTy [167] is a frequently-cited variation on the classic BERT transformer [168] which was trained on 558M natural antibody sequences to learn the "antibody language" and produce rich embeddings of antibodies. AntiBERTa [169] is another pLM based on the RoBERTa transformer architecture [170] which was trained on 42M heavy chains and 15M light chains to produce rich representations of B cell receptors. Furthermore, these model can be fine-tuned for paratope generation. Other antibody language models include IgBert and IgT5 [171] which are variants of the BERT and T5 transformer [172] architectures and AbLang [173] which also is a RoBERTa variant.

Several groups have proposed generative sequence-based models for the design of antibodies. Among those, we find the Immunoglobulin Language Model (IgLM) [174] which is based on the GPT-2 architecture [175]. IgLM was trained on 558M natural antibody sequences from the Observed antibody space database [176] to infill fragments of masked residues in antibody sequences. By doing so, it learned to generate human-like antibodies. One application of IgLM proposed by the authors is the generation of CDR loop variants, which can be assembled into a library and screened as part of an antibody optimization pipeline.

Recently, Hie *et al.* [177] used the ESM-1b pLM [178] and the ESM-1v pLM ensemble [179] to optimize or "mature" seven antibodies against coronavirus, ebolavirus and influenza A virus. They were able to produce antibodies with improved neutralization activity after only 2 rounds of laboratory evaluation and evolution. Interestingly, they suggest that using models trained on all protein sequences as opposed to antibody sequences only (*e.g.* AntiBERTy/AntiBERTa) is advantageous as these models learned more general rules of evolution. To optimize the antibody sequences, they select from mutations in the antibody sequences predicted by the pLMs to be most likely to occur as part of natural evolution.

## Summary and future trends

Until recently, machine learning models such as SVMs and CNNs trained on human-engineered features constituted the majority of sequence-based PPI predictors. However, there is currently a clear trend towards pLMs-generated protein representations. This paradigm shift towards language modeling of biopolymers now expands beyond protein classification and design and well into genomics [180], [181], [182] and transcriptomics [183], [184].

Progress in PPI prediction remains slow, nonetheless. A panoply of models exploiting advances in deep learning have been proposed within the last decade, but similarity-based methods such as PIPE and SPRINT remain competitive in rigorous benchmarks, even if they predate them [28]. Progress is hindered by factors such as flawed methodology. For example, a surprisingly large number of authors fail to account for class imbalance and evaluate their models on balanced test sets. Dunham *et al.* found that most published methods experience a dramatic drop in performance when applied to realistic and imbalanced test sets [77]. In addition, the notion of pair difficulty (C1/C2/C3) introduced by Park and Marcotte [76] which we discussed is also ignored by most groups. This causes "data leakage", *i.e.* the sharing of information from the training set to the test set, which should be completely independent. Data leakage thus also leads to performance overestimates and has been found to be widespread issue [28].

Issues surrounding model evaluation could be mitigated through the creation and maintenance of standardized benchmark test sets of carefully curated PPIs which have not been deposited in public databases, as is done in the protein structure prediction community with the Critical Assessment of Structure Prediction (CASP). Standardized benchmark datasets are also customary in many machine learning subdisciplines such as natural language processing and computer vision.

In contrast with structure-based methods, sequence-based PPI predictors do not consider proteins as static structures and do not rely on potentially inaccurate predicted structures or scarce high quality experimentally determined protein structures. Not only do they facilitate the identification of actionable drug targets within protein interaction networks, but they can also be used to design therapeutic biologics such as peptide binders and antibodies. There is a growing interest for biologics as a treatment modality, considering that they were found to be more likely to succeed in clinical trials than small molecules [185]. It will be interesting to see whether academia and the biotechnology industry decide to invest in and capitalize on the promise of sequence-based PPI prediction.

# References

[1] H. Elhabashy, F. Merino, V. Alva, O. Kohlbacher, and A. N. Lupas, "Exploring Protein-Protein Interactions at the Proteome Level," *Structure*, vol. 30, no. 4, pp. 462–475, Apr. 2022, doi: 10.1016/j.str.2022.02.004.

[2] I. M. A. Nooren and J. M. Thornton, "Diversity of Protein–Protein Interactions," *The EMBO Journal*, vol. 22, no. 14, pp. 3486–3492, 2003, doi: 10.1093/emboj/cdg359.

[3] P. Friedhoff, P. Li, and J. Gotthardt, "Protein-Protein Interactions in DNA Mismatch Repair," *DNA Repair*, vol. 38, pp. 50–57, Feb. 2016, doi: 10.1016/j.dnarep.2015.11.013.

[4] D. A. Guarracino, B. N. Bullock, and P. S. Arora, "Protein-Protein Interactions in Transcription: A Fertile Ground for Helix Mimetics," *Biopolymers*, vol. 95, no. 1, pp. 1–7, Jan. 2011, doi: 10.1002/bip.21546.

[5] X. Jia, X. He, C. Huang, J. Li, Z. Dong, and K. Liu, "Protein Translation: Biological Processes and Therapeutic Strategies for Human Diseases," *Signal Transduction and Targeted Therapy*, vol. 9, no. 1, p. 44, Feb. 2024, doi: 10.1038/s41392-024-01749-9.

[6] J. Westermarck, J. Ivaska, and G. L. Corthals, "Identification of Protein Interactions Involved in Cellular Signaling," *Molecular & Cellular Proteomics : MCP*, vol. 12, no. 7, pp. 1752–1763, Jul. 2013, doi: 10.1074/mcp.R113.027771.

[7] J. Buchner, "Molecular Chaperones and Protein Quality Control: An Introduction to the JBC Reviews Thematic Series," *The Journal of Biological Chemistry*, vol. 294, no. 6, pp. 2074–2075, Feb. 2019, doi: 10.1074/jbc.REV118.006739.

[8] M. W. Gonzalez and M. G. Kann, "Chapter 4: Protein Interactions and Disease," *PLoS Computational Biology*, vol. 8, no. 12, p. e1002819, Dec. 2012, doi: 10.1371/journal.pcbi.1002819.

[9] F. Cheng *et al.*, "Comprehensive Characterization of Protein–Protein Interactions Perturbed by Disease Mutations," *Nature Genetics*, vol. 53, no. 3, pp. 342–353, Mar. 2021, doi: 10.1038/s41588-020-00774-y.

[10] J. F. Greenblatt, B. M. Alberts, and N. J. Krogan, "Discovery and Significance of Protein-Protein Interactions in Health and Disease," *Cell*, vol. 187, no. 23, pp. 6501–6517, Nov. 2024, doi: 10.1016/j.cell.2024.10.038.

[11] D. S. Knopman *et al.*, "Alzheimer Disease," *Nature Reviews Disease Primers*, vol. 7, no. 1, p. 33, May 2021, doi: 10.1038/s41572-021-00269-y.

[12] J. Janssens and C. Van Broeckhoven, "Pathological Mechanisms Underlying TDP-43 Driven Neurodegeneration in FTLD–ALS Spectrum Disorders," *Human Molecular Genetics*, vol. 22, no. R1, pp. R77–R87, Oct. 2013, doi: 10.1093/hmg/ddt349.

[13] B. R. Bloem, M. S. Okun, and C. Klein, "Parkinson's Disease," *The Lancet*, vol. 397, no. 10291, pp. 2284–2303, Jun. 2021, doi: 10.1016/s0140-6736(21)00218-x.

[14] G. P. Bates *et al.*, "Huntington Disease," *Nature Reviews Disease Primers*, vol. 1, no. 1, p. 15005, Apr. 2015, doi: 10.1038/nrdp.2015.5.

[15] Y. Iwasaki, "Creutzfeldt-Jakob Disease," *Neuropathology*, vol. 37, no. 2, pp. 174–188, 2017, doi: 10.1111/neup.12355.

[16] L. Huang, Z. Guo, F. Wang, and L. Fu, "KRAS Mutation: From Undruggable to Druggable in Cancer," *Signal Transduction and Targeted Therapy*, vol. 6, no. 1, p. 386, Nov. 2021, doi: 10.1038/s41392-021-00780-4.

[17] M. S. Tabar, C. Parsania, H. Chen, X.-D. Su, C. G. Bailey, and J. E. J. Rasko, "Illuminating the Dark Protein-Protein Interactome," *Cell Reports Methods*, vol. 2, no. 8, Aug. 2022, doi: [10.1016/j.crmeth.2022.100275](10.1016/j.crmeth.2022.100275).

[18] M. Kim *et al.*, "A Protein Interaction Landscape of Breast Cancer," *Science*, vol. 374, no. 6563, p. eabf3066, Oct. 2021, doi: [10.1126/science.abf3066](10.1126/science.abf3066).

[19] H. Fu, X. Mo, and A. A. Ivanov, "Decoding the Functional Impact of the Cancer Genome through Protein–Protein Interactions," *Nature Reviews Cancer*, vol. 25, no. 3, pp. 189–208, Mar. 2025, doi: [10.1038/s41568-024-00784-6](10.1038/s41568-024-00784-6).

[20] W. H. Dunham, M. Mullin, and A.-C. Gingras, "Affinity-purification Coupled to Mass Spectrometry: Basic Principles and Strategies," *PROTEOMICS*, vol. 12, no. 10, pp. 1576–1590, May 2012, doi: [10.1002/pmic.201100523](10.1002/pmic.201100523).

[21] S. S. Sidhu, W. J. Fairbrother, and K. Deshayes, "Exploring Protein–Protein Interactions with Phage Display," *ChemBioChem*, vol. 4, no. 1, pp. 14–25, Jan. 2003, doi: [10.1002/cbic.200390008](10.1002/cbic.200390008).

[22] M. Zhou, Q. Li, and R. Wang, "Current Experimental Methods for Characterizing Protein–Protein Interactions," *ChemMedChem*, vol. 11, no. 8, pp. 738–756, Apr. 2016, doi: [10.1002/cmdc.201500495](10.1002/cmdc.201500495).

[23] A. Brückner, C. Polge, N. Lentze, D. Auerbach, and U. Schlattner, "Yeast Two-Hybrid, a Powerful Tool for Systems Biology," *International Journal of Molecular Sciences*, vol. 10, no. 6, pp. 2763–2788, Jun. 2009, doi: [10.3390/ijms10062763](10.3390/ijms10062763).

[24] S. Akbarzadeh, Ö. Coşkun, and B. Güncer, "Studying Protein–Protein Interactions: Latest and Most Popular Approaches," *Journal of Structural Biology*, vol. 216, no. 4, p. 108118, Dec. 2024, doi: [10.1016/j.jsb.2024.108118](10.1016/j.jsb.2024.108118).

[25] S. Pitre *et al.*, "Global Investigation of Protein–Protein Interactions in Yeast Saccharomyces Cerevisiae Using Re-Occurring Short Polypeptide Sequences," *Nucleic Acids Research*, vol. 36, no. 13, pp. 4286–4294, Aug. 2008, doi: [10.1093/nar/gkn390](10.1093/nar/gkn390).

[26] Y. Li and L. Ilie, "SPRINT: Ultrafast Protein-Protein Interaction Prediction of the Entire Human Interactome," *BMC Bioinformatics*, vol. 18, no. 1, p. 485, Nov. 2017, doi: [10.1186/s12859-017-1871-x](10.1186/s12859-017-1871-x).

[27] K. Dick *et al.*, "PIPE4: Fast PPI Predictor for Comprehensive Inter- and Cross-Species Interactomes," *Scientific Reports*, vol. 10, no. 1, p. 1390, Dec. 2020, doi: [10.1038/s41598-019-56895-w](10.1038/s41598-019-56895-w).

[28] J. Bernett, D. B. Blumenthal, and M. List, "Cracking the Black Box of Deep Sequence-Based Protein–Protein Interaction Prediction," *Briefings in Bioinformatics*, vol. 25, no. 2, p. bbae76, Mar. 2024, doi: [10.1093/bib/bbae076](10.1093/bib/bbae076).

[29] S. A. Andrei *et al.*, "Stabilization of Protein-Protein Interactions in Drug Discovery," *Expert Opinion on Drug Discovery*, vol. 12, no. 9, pp. 925–940, Sep. 2017, doi: [10.1080/17460441.2017.1346608](10.1080/17460441.2017.1346608).

[30] S. J. Y. Macalino, S. Basith, N. A. B. Clavio, H. Chang, S. Kang, and S. Choi, "Evolution of In Silico Strategies for Protein-Protein Interaction Drug Discovery," *Molecules*, vol. 23, no. 8, p. 1963, Aug. 2018, doi: [10.3390/molecules23081963](10.3390/molecules23081963).

[31] X. Wang, D. Ni, Y. Liu, and S. Lu, "Rational Design of Peptide-Based Inhibitors Disrupting Protein-Protein Interactions," *Frontiers in Chemistry*, vol. 9, May 2021, doi: [10.3389/fchem.2021.682675](10.3389/fchem.2021.682675).

[32] P. W. Rose *et al.*, "The RCSB Protein Data Bank: Integrative View of Protein, Gene and 3D Structural Information," *Nucleic Acids Research*, vol. 45, no. D1, pp. D271–D281, Jan. 2017, doi: [10.1093/nar/gkw1000](10.1093/nar/gkw1000).

[33] R. Oughtred *et al.*, "The BioGRID Database: A Comprehensive Biomedical Resource of Curated Protein, Genetic, and Chemical Interactions," *Protein Science*, vol. 30, no. 1, pp. 187–200, 2021, doi: [10.1002/pro.3978](10.1002/pro.3978).

[34] J. Jumper *et al.*, "Highly Accurate Protein Structure Prediction with AlphaFold," *Nature*, vol. 596, no. 7873, pp. 583–589, Aug. 2021, doi: 10.1038/s41586-021-03819-2.

[35] J. Abramson *et al.*, "Accurate Structure Prediction of Biomolecular Interactions with AlphaFold 3," *Nature*, vol. 630, no. 8016, pp. 493–500, Jun. 2024, doi: 10.1038/s41586-024-07487-w.

[36] Z. Lin *et al.*, "Evolutionary-Scale Prediction of Atomic-Level Protein Structure with a Language Model," *Science*, vol. 379, no. 6637, pp. 1123–1130, Mar. 2023, doi: 10.1126/science.ade2574.

[37] Chai Discovery *et al.*, "Chai-1: Decoding the Molecular Interactions of Life." bioRxiv, Oct. 2024. doi: 10.1101/2024.10.10.615955.

[38] J. Wohlwend *et al.*, "Boltz-1 Democratizing Biomolecular Interaction Modeling." bioRxiv, Nov. 2024. doi: 10.1101/2024.11.19.624167.

[39] S. Passaro *et al.*, "Boltz-2: Towards Accurate and Efficient Binding Affinity Prediction." bioRxiv, Jun. 2025. doi: 10.1101/2025.06.14.659707.

[40] T. C. Terwilliger *et al.*, "AlphaFold Predictions Are Valuable Hypotheses and Accelerate but Do Not Replace Experimental Structure Determination," *Nature Methods*, vol. 21, no. 1, pp. 110–116, Jan. 2024, doi: 10.1038/s41592-023-02087-4.

[41] J. Verburgt, Z. Zhang, and D. Kihara, "Multi-Level Analysis of Intrinsically Disordered Protein Docking Methods," *Methods*, vol. 204, pp. 55–63, Aug. 2022, doi: 10.1016/j.ymeth.2022.05.006.

[42] G. Kibar and M. Vingron, "Prediction of Protein–Protein Interactions Using Sequences of Intrinsically Disordered Regions," *Proteins: Structure, Function, and Bioinformatics*, vol. 91, no. 7, pp. 980–990, 2023, doi: 10.1002/prot.26486.

[43] C. Y. Lee *et al.*, "Systematic Discovery of Protein Interaction Interfaces Using AlphaFold and Experimental Validation," *Molecular Systems Biology*, vol. 20, no. 2, pp. 75–97, Feb. 2024, doi: 10.1038/s44320-023-00005-6.

[44] T. Orand and M. R. Jensen, "Binding Mechanisms of Intrinsically Disordered Proteins: Insights from Experimental Studies and Structural Predictions," *Current Opinion in Structural Biology*, vol. 90, p. 102958, Feb. 2025, doi: 10.1016/j.sbi.2024.102958.

[45] F. Luppino, S. Lenz, C. F. W. Chow, and A. Toth-Petroczy, "Deep Learning Tools Predict Variants in Disordered Regions with Lower Sensitivity," *BMC Genomics*, vol. 26, no. 1, p. 367, Apr. 2025, doi: 10.1186/s12864-025-11534-9.

[46] R. Yuan, J. Zhang, J. Zhou, and Q. Cong, "Recent Progress and Future Challenges in Structure-Based Protein-Protein Interaction Prediction," *Molecular Therapy*, vol. 33, no. 5, pp. 2252–2268, May 2025, doi: 10.1016/j.ymthe.2025.04.003.

[47] N. Raisinghani, V. Parikh, B. Foley, and G. Verkhivker, "Assessing Structures and Conformational Ensembles of Apo and Holo Protein States Using Randomized Alanine Sequence Scanning Combined with Shallow Subsampling in AlphaFold2 : Insights and Lessons from Predictions of Functional Allosteric Conformations." Cold Spring Harbor Laboratory, Nov. 2024. doi: 10.1101/2024.11.04.621947.

[48] T. Mitchell, *Machine Learning*. in McGraw-Hill Series in Computer Science. New York, NY: McGraw-Hill Professional, 1997.

[49] K. Yugandhar and M. M. Gromiha, "Protein–Protein Binding Affinity Prediction from Amino Acid Sequence," *Bioinformatics*, vol. 30, no. 24, pp. 3583–3589, Dec. 2014, doi: 10.1093/bioinformatics/btu580.

[50] W. A. Abbasi, A. Yaseen, F. U. Hassan, S. Andleeb, and F. U. A. A. Minhas, "ISLAND: In-Silico Proteins Binding Affinity Prediction Using Sequence Information," *BioData Mining*, vol. 13, no. 1, pp. 1–13, Dec. 2020, doi: 10.1186/s13040-020-00231-w.

[51] Z. Guo and R. Yamaguchi, "Machine Learning Methods for Protein-Protein Binding Affinity Prediction in Protein Design," *Frontiers in Bioinformatics*, vol. 2, Dec. 2022, doi: 10.3389/fbinf.2022.1065703.

[52] S. Romero-Molina *et al.*, "PPI-Affinity: A Web Tool for the Prediction and Optimization of Protein–Peptide and Protein–Protein Binding Affinity," *Journal of Proteome Research*, vol. 21, no. 8, pp. 1829–1841, Aug. 2022, doi: 10.1021/acs.jproteome.2c00020.

[53] Y. Ofran and B. Rost, "Predicted Protein–Protein Interaction Sites from Local Sequence Information," *FEBS Letters*, vol. 544, no. 1, pp. 236–239, Jun. 2003, doi: 10.1016/S0014-5793(03)00456-3.

[54] I. Ezkurdia, L. Bartoli, P. Fariselli, R. Casadio, A. Valencia, and M. L. Tress, "Progress and Challenges in Predicting Protein–Protein Interaction Sites," *Briefings in Bioinformatics*, vol. 10, no. 3, pp. 233–246, May 2009, doi: 10.1093/bib/bbp021.

[55] J. Zhang and L. Kurgan, "Review and Comparative Assessment of Sequence-Based Predictors of Protein-Binding Residues," *Briefings in Bioinformatics*, vol. 19, no. 5, pp. 821–837, Sep. 2018, doi: 10.1093/bib/bbx022.

[56] Y. Li, G. B. Golding, and L. Ilie, "DELPHI: Accurate Deep Ensemble Model for Protein Interaction Sites Prediction," *Bioinformatics*, vol. 37, no. 7, pp. 896–904, May 2021, doi: 10.1093/bioinformatics/btaa750.

[57] Y. Guo, L. Yu, Z. Wen, and M. Li, "Using Support Vector Machine Combined with Auto Covariance to Predict Protein–Protein Interactions from Protein Sequences," *Nucleic Acids Research*, vol. 36, no. 9, pp. 3025–3030, May 2008, doi: 10.1093/nar/gkn159.

[58] A. Ben-Hur and W. S. Noble, "Choosing Negative Examples for the Prediction of Protein-Protein Interactions," *BMC Bioinformatics*, vol. 7, no. Suppl1, p. S2, Mar. 2006, doi: 10.1186/1471-2105-7-S1-S2.

[59] S. Romero-Molina, Y. B. Ruiz-Blanco, M. Harms, J. Münch, and E. Sanchez-Garcia, "PPI-Detect: A Support Vector Machine Model for Sequence-Based Prediction of Protein-Protein Interactions: PPI-Detect: A Support Vector Machine Model for Sequence-Based Prediction of Protein-Protein Interactions," *Journal of Computational Chemistry*, vol. 40, no. 11, pp. 1233–1242, Apr. 2019, doi: 10.1002/jcc.25780.

[60] P. Smialowski *et al.*, "The Negatome Database: A Reference Set of Non-Interacting Protein Pairs," *Nucleic Acids Research*, vol. 38, no. suppl_1, pp. D540–D544, Jan. 2010, doi: 10.1093/nar/gkp1026.

[61] P. Blohm *et al.*, "Negatome 2.0: A Database of Non-Interacting Proteins Derived by Literature Mining, Manual Annotation and Protein Structure Analysis," *Nucleic Acids Research*, vol. 42, no. Database issue, pp. D396–D400, Jan. 2014, doi: 10.1093/nar/gkt1079.

[62] The UniProt Consortium, "UniProt: The Universal Protein Knowledgebase in 2025," *Nucleic Acids Research*, vol. 53, no. D1, pp. D609–D617, Jan. 2025, doi: 10.1093/nar/gkae1010.

[63] D. Szklarczyk *et al.*, "The STRING Database in 2023: Protein-Protein Association Networks and Functional Enrichment Analyses for Any Sequenced Genome of Interest," *Nucleic Acids Research*, vol. 51, no. D1, pp. D638–D646, Jan. 2023, doi: 10.1093/nar/gkac1000.

[64] N. del~Toro *et al.*, "The IntAct Database: Efficient Access to Fine-Grained Molecular Interaction Data," *Nucleic Acids Research*, vol. 50, no. D1, pp. D648–D653, Jan. 2022, doi: 10.1093/nar/gkab1006.

[65] L. Licata *et al.*, "MINT, the Molecular Interaction Database: 2012 Update," *Nucleic Acids Research*, vol. 40, no. Database issue, pp. D857–861, Jan. 2012, doi: 10.1093/nar/gkr930.

[66] P. Martins *et al.*, "Propedia v2.3: A Novel Representation Approach for the Peptide-Protein Interaction Database Using Graph-Based Structural Signatures," *Frontiers in Bioinformatics*, vol. 3, Feb. 2023, doi: 10.3389/fbinf.2023.1103103.

[67] K. Sikic and O. Carugo, "Protein Sequence Redundancy Reduction: Comparison of Various Method," *Bioinformation*, vol. 5, no. 6, pp. 234–239, Nov. 2010, doi: 10.6026/97320630005234.

[68] L. Fu, B. Niu, Z. Zhu, S. Wu, and W. Li, "CD-HIT: Accelerated for Clustering the next-Generation Sequencing Data," *Bioinformatics*, vol. 28, no. 23, pp. 3150–3152, Dec. 2012, doi: 10.1093/bioinformatics/bts565.

[69] M. Chen *et al.*, "Multifaceted Protein–Protein Interaction Prediction Based on Siamese Residual RCNN," *Bioinformatics*, vol. 35, no. 14, pp. i305–i314, Jul. 2019, doi: 10.1093/bioinformatics/btz328.

[70] S. Sledzieski, R. Singh, L. Cowen, and B. Berger, "D-SCRIPT Translates Genome to Phenome with Sequence-Based, Structure-Aware, Genome-Scale Predictions of Protein-Protein Interactions," *Cell Systems*, vol. 12, no. 10, pp. 969–982, Oct. 2021, doi: 10.1016/j.cels.2021.08.010.

[71] B. Yu, C. Chen, H. Zhou, B. Liu, and Q. Ma, "Gradient," *Genomics, Proteomics & Bioinformatics*, vol. 18, no. 5, pp. 582–592, Oct. 2020, doi: 10.1016/j.gpb.2021.01.001.

[72] M. Steinegger and J. Söding, "MMseqs2 Enables Sensitive Protein Sequence Searching for the Analysis of Massive Data Sets," *Nature Biotechnology*, vol. 35, no. 11, pp. 1026–1028, Nov. 2017, doi: 10.1038/nbt.3988.

[73] C. Chen *et al.*, "Improving Protein-Protein Interactions Prediction Accuracy Using XGBoost Feature Selection and Stacked Ensemble Classifier," *Computers in Biology and Medicine*, vol. 123, p. 103899, Aug. 2020, doi: 10.1016/j.compbiomed.2020.103899.

[74] D. Liu *et al.*, "PLM-interact: Extending Protein Language Models to Predict Protein-Protein Interactions." Nov. 2024. doi: 10.1101/2024.11.05.622169.

[75] X. Zheng *et al.*, "PRING: Rethinking Protein-Protein Interaction Prediction from Pairs to Graphs," no. arXiv:2507.05101. arXiv, Jul. 2025. doi: 10.48550/arXiv.2507.05101.

[76] Y. Park and E. M. Marcotte, "Flaws in Evaluation Schemes for Pair-Input Computational Predictions," *Nature Methods*, vol. 9, no. 12, pp. 1134–1136, Dec. 2012, doi: 10.1038/nmeth.2259.

[77] B. Dunham and M. K. Ganapathiraju, "Benchmark Evaluation of Protein–Protein Interaction Prediction Algorithms," *Molecules*, vol. 27, no. 1, p. 41, Dec. 2021, doi: 10.3390/molecules27010041.

[78] R. O. Duda, D. G. Stork, and P. E. Hart, *Pattern Classification*, 2nd ed. New York: Wiley, 2001.

[79] C. M. Bishop, *Pattern Recognition and Machine Learning*. in Information Science and Statistics. New York: Springer, 2006.

[80] I. Goodfellow, Y. Bengio, and A. Courville, *Deep Learning*. in Adaptive Computation and Machine Learning. Cambridge, Massachusetts: The MIT Press, 2016.

[81] K. Dick, A. Chopra, K. K. Biggar, and J. R. Green, "Multi-Schema Computational Prediction of the Comprehensive SARS-CoV-2 vs. Human Interactome," *PeerJ*, vol. 9, p. e11117, Apr. 2021, doi: 10.7717/peerj.11117.

[82] X. Yang, S. Yang, Q. Li, S. Wuchty, and Z. Zhang, "Prediction of Human-Virus Protein-Protein Interactions through a Sequence Embedding-Based Machine Learning Method," *Computational and Structural Biotechnology Journal*, vol. 18, pp. 153–161, Jan. 2020, doi: 10.1016/j.csbj.2019.12.005.

[83] S. Tsukiyama, M. M. Hasan, S. Fujii, and H. Kurata, "LSTM-PHV: Prediction of Human-Virus Protein–Protein Interactions by LSTM with Word2vec," *Briefings in Bioinformatics*, vol. 22, no. 6, p. bbab228, Nov. 2021, doi: 10.1093/bib/bbab228.

[84] X. Yang, S. Yang, X. Lian, S. Wuchty, and Z. Zhang, "Transfer Learning via Multi-Scale Convolutional Neural Layers for Human–Virus Protein–Protein Interaction Prediction," *Bioinformatics*, vol. 37, no. 24, pp. 4771–4778, Dec. 2021, doi: 10.1093/bioinformatics/btab533.

[85]  T. N. Dong, G. Brogden, G. Gerold, and M. Khosla, "A Multitask Transfer Learning Framework for the Prediction of Virus-Human Protein–Protein Interactions," *BMC Bioinformatics*, vol. 22, no. 1, pp. 1–24, Dec. 2021, doi: 10.1186/s12859-021-04484-y.

[86]  N. Nissan *et al.*, "Large-Scale Data Mining Pipeline for Identifying Novel Soybean Genes Involved in Resistance against the Soybean Cyst Nematode," *Frontiers in Bioinformatics*, vol. 3, Jun. 2023, doi: 10.3389/fbinf.2023.1199675.

[87]  B. Barnes *et al.*, "Predicting Novel Protein-Protein Interactions between the HIV-1 Virus and Homo Sapiens," in *2016 IEEE EMBS International Student Conference (ISC)*, May 2016, pp. 1–4. doi: 10.1109/EMBSISC.2016.7508598.

[88]  R. Singh, K. Devkota, S. Sledzieski, B. Berger, and L. Cowen, "Topsy-Turvy: Integrating a Global View into Sequence-Based PPI Prediction," *Bioinformatics*, vol. 38, no. Supplement_1, pp. i264–i272, Jun. 2022, doi: 10.1093/bioinformatics/btac258.

[89]  J. Szymborski and A. Emad, "INTREPPPID—an Orthologue-Informed Quintuplet Network for Cross-Species Prediction of Protein–Protein Interaction," *Briefings in Bioinformatics*, vol. 25, no. 5, p. bbae405, Sep. 2024, doi: 10.1093/bib/bbae405.

[90]  K. Volzhenin, L. Bittner, and A. Carbone, "SENSE-PPI Reconstructs Interactomes within, across, and between Species at the Genome Scale," *iScience*, vol. 27, no. 7, p. 110371, Jul. 2024, doi: 10.1016/j.isci.2024.110371.

[91]  V. Ajila *et al.*, "Species-Specific microRNA Discovery and Target Prediction in the Soybean Cyst Nematode," *Scientific Reports*, vol. 13, no. 1, p. 17657, Oct. 2023, doi: 10.1038/s41598-023-44469-w.

[92]  K. K. Biggar *et al.*, "Proteome-Wide Prediction of Lysine Methylation Leads to Identification of H2BK43 Methylation and Outlines the Potential Methyllysine Proteome," *Cell Reports*, vol. 32, no. 2, p. 107896, Jul. 2020, doi: 10.1016/j.celrep.2020.107896.

[93]  G. Wang, I. I. Vaisman, and M. L. van Hoek, "Machine Learning Prediction of Antimicrobial Peptides," *Methods in molecular biology (Clifton, N.J.)*, vol. 2405, pp. 1–37, 2022, doi: 10.1007/978-1-0716-1855-4_1.

[94]  D. Brzezinski, L. L. Minku, T. Pewinski, J. Stefanowski, and A. Szumaczuk, "The Impact of Data Difficulty Factors on Classification of Imbalanced and Concept Drifting Data Streams," *Knowledge and Information Systems*, vol. 63, no. 6, pp. 1429–1469, Jun. 2021, doi: 10.1007/s10115-021-01560-w.

[95]  E. Richardson, R. Trevizani, J. A. Greenbaum, H. Carter, M. Nielsen, and B. Peters, "The Receiver Operating Characteristic Curve Accurately Assesses Imbalanced Datasets," *Patterns*, vol. 5, no. 6, p. 100994, Jun. 2024, doi: 10.1016/j.patter.2024.100994.

[96]  M. Langote, N. Zade, and S. Gundewar, "Addressing Data Imbalance in Machine Learning: Challenges and Approaches," in *2025 6th International Conference on Mobile Computing and Sustainable Informatics (ICMCSI)*, Jan. 2025, pp. 1745–1749. doi: 10.1109/ICMCSI64620.2025.10883059.

[97]  K. Dick and J. R. Green, "Reciprocal Perspective for Improved Protein-Protein Interaction Prediction," *Scientific Reports*, vol. 8, no. 1, p. 11694, 2018, doi: 10.1038/s41598-018-30044-1.

[98]  K. Venkatesan *et al.*, "An Empirical Framework for Binary Interactome Mapping," *Nature Methods*, vol. 6, no. 1, pp. 83–90, Jan. 2009, doi: 10.1038/nmeth.1280.

[99]  M. P. H. Stumpf *et al.*, "Estimating the Size of the Human Interactome," *Proceedings of the National Academy of Sciences*, vol. 105, no. 19, pp. 6959–6964, May 2008, doi: 10.1073/pnas.0708078105.

[100]  K.-C. Chou, "Prediction of Protein Cellular Attributes Using Pseudo-Amino Acid Composition," *Proteins: Structure, Function, and Bioinformatics*, vol. 43, no. 3, pp. 246–255, 2001, doi: 10.1002/prot.1035.

[101] J. Shen *et al.*, "Predicting Protein–Protein Interactions Based Only on Sequences Information," *Proceedings of the National Academy of Sciences*, vol. 104, no. 11, pp. 4337–4341, Mar. 2007, doi: 10.1073/pnas.0607879104.

[102] G. Govindan and A. S. Nair, "Composition, Transition and Distribution (CTD) — A Dynamic Feature for Predictions Based on Hierarchical Structure of Cellular Sorting," in *2011 Annual IEEE India Conference*, Dec. 2011, pp. 1–6. doi: 10.1109/INDCON.2011.6139332.

[103] S. F. Altschul *et al.*, "Gapped BLAST and PSI-BLAST: A New Generation of Protein Database Search Programs," *Nucleic Acids Research*, vol. 25, no. 17, pp. 3389–3402, Sep. 1997, doi: 10.1093/nar/25.17.3389.

[104] Y. B. Ruiz-Blanco, W. Paz, J. Green, and Y. Marrero-Ponce, "ProtDCal: A Program to Compute General-Purpose-Numerical Descriptors for Sequences and 3D-structures of Proteins," *BMC Bioinformatics*, vol. 16, no. 1, p. 162, May 2015, doi: 10.1186/s12859-015-0586-0.

[105] T. Mikolov, K. Chen, G. Corrado, and J. Dean, "Efficient Estimation of Word Representations in Vector Space," no. arXiv:1301.3781. arXiv, Sep. 2013. doi: 10.48550/arXiv.1301.3781.

[106] Y. Yao, X. Du, Y. Diao, and H. Zhu, "An Integration of Deep Learning with Feature Embedding for Protein-Protein Interaction Prediction," *PeerJ*, vol. 7, p. e7126, 2019, doi: 10.7717/peerj.7126.

[107] S. Hashemifar, B. Neyshabur, A. A. Khan, and J. Xu, "Predicting Protein–Protein Interactions through Sequence-Based Deep Learning," *Bioinformatics*, vol. 34, no. 17, pp. i802–i810, Sep. 2018, doi: 10.1093/bioinformatics/bty573.

[108] X. Hu, C. Feng, Y. Zhou, A. Harrison, and M. Chen, "DeepTrio: A Ternary Prediction System for Protein–Protein Interaction Using Mask Multiple Parallel Convolutional Neural Networks," *Bioinformatics*, vol. 38, no. 3, pp. 694–702, Jan. 2022, doi: 10.1093/bioinformatics/btab737.

[109] F. Soleymani, E. Paquet, H. L. Viktor, W. Michalowski, and D. Spinello, "ProtInteract: A Deep Learning Framework for Predicting Protein–Protein Interactions," *Computational and Structural Biotechnology Journal*, vol. 21, pp. 1324–1348, Jan. 2023, doi: 10.1016/j.csbj.2023.01.028.

[110] J. Hu, Z. Li, B. Rao, M. A. Thafar, and M. Arif, "Improving Protein-Protein Interaction Prediction Using Protein Language Model and Protein Network Features," *Analytical Biochemistry*, vol. 693, p. 115550, Oct. 2024, doi: 10.1016/j.ab.2024.115550.

[111] T. H. Dang and T. A. Vu, "xCAPT5: Protein–Protein Interaction Prediction Using Deep and Wide Multi-Kernel Pooling Convolutional Neural Networks with Protein Language Model," *BMC Bioinformatics*, vol. 25, no. 1, pp. 1–20, Dec. 2024, doi: 10.1186/s12859-024-05725-6.

[112] F.-E. Eid, M. ElHefnawi, and L. S. Heath, "DeNovo: Virus-Host Sequence-Based Protein–Protein Interaction Prediction," *Bioinformatics*, vol. 32, no. 8, pp. 1144–1150, Apr. 2016, doi: 10.1093/bioinformatics/btv737.

[113] H. Gao, C. Chen, S. Li, C. Wang, W. Zhou, and B. Yu, "Prediction of Protein-Protein Interactions Based on Ensemble Residual Convolutional Neural Network," *Computers in Biology and Medicine*, vol. 152, p. 106471, Jan. 2023, doi: 10.1016/j.compbiomed.2022.106471.

[114] X. Li, P. Han, G. Wang, W. Chen, S. Wang, and T. Song, "SDNN-PPI: Self-Attention with Deep Neural Network Effect on Protein-Protein Interaction Prediction," *BMC Genomics*, vol. 23, no. 1, pp. 1–14, Dec. 2022, doi: 10.1186/s12864-022-08687-2.

[115] A. Vaswani *et al.*, "Attention Is All You Need." arXiv, 2017. doi: 10.48550/ARXIV.1706.03762.

[116] OpenAI *et al.*, "GPT-4 Technical Report," no. arXiv:2303.08774. arXiv, Mar. 2024. doi: 10.48550/arXiv.2303.08774.

[117] G. Team *et al.*, "Gemini 1.5: Unlocking Multimodal Understanding across Millions of Tokens of Context," no. arXiv:2403.05530. arXiv, Dec. 2024. doi: 10.48550/arXiv.2403.05530.

[118] A. Krogh, M. Brown, I. S. Mian, K. Sjölander, and D. Haussler, "Hidden Markov Models in Computational Biology: Applications to Protein Modeling," *Journal of Molecular Biology*, vol. 235, no. 5, pp. 1501–1531, Feb. 1994, doi: 10.1006/jmbi.1994.1104.

[119] P. L. Martelli, P. Fariselli, A. Krogh, and R. Casadio, "A Sequence-Profile-Based HMM for Predicting and Discriminating $b\eta$ Barrel Membrane Proteins", *Bioinformatics*, vol. 18, no. suppl_1, pp. S46–S53, Jul. 2002, doi: 10.1093/bioinformatics/18.suppl_1.s46.

[120] J. Söding, "Protein Homology Detection by HMM–HMM Comparison," *Bioinformatics*, vol. 21, no. 7, pp. 951–960, Apr. 2005, doi: 10.1093/bioinformatics/bti125.

[121] T. Bepler and B. Berger, "Learning the Protein Language: Evolution, Structure, and Function," *Cell Systems*, vol. 12, no. 6, pp. 654–669, Jun. 2021, doi: 10.1016/j.cels.2021.05.017.

[122] B. E. Suzek, Y. Wang, H. Huang, P. B. McGarvey, and C. H. Wu, "UniRef Clusters: A Comprehensive and Scalable Alternative for Improving Sequence Similarity Searches," *Bioinformatics*, vol. 31, no. 6, pp. 926–932, Mar. 2015, doi: 10.1093/bioinformatics/btu739.

[123] M. Steinegger and J. Söding, "Clustering Huge Protein Sequence Sets in Linear Time," *Nature Communications*, vol. 9, no. 1, p. 2542, Jun. 2018, doi: 10.1038/s41467-018-04964-5.

[124] D. Medina-Ortiz *et al.*, "Protein Language Models and Machine Learning Facilitate the Identification of Antimicrobial Peptides," *International Journal of Molecular Sciences*, vol. 25, no. 16, p. 8851, Aug. 2024, doi: 10.3390/ijms25168851.

[125] L. Zhang *et al.*, "Leveraging Protein Language Models for Robust Antimicrobial Peptide Detection," *Methods*, vol. 238, pp. 19–26, Jun. 2025, doi: 10.1016/j.ymeth.2025.03.002.

[126] Y. S. Ko, J. Parkinson, C. Liu, and W. Wang, "TUnA: An Uncertainty-Aware Transformer Model for Sequence-Based Protein–Protein Interaction Prediction," *Briefings in Bioinformatics*, vol. 25, no. 5, p. bbae359, Sep. 2024, doi: 10.1093/bib/bbae359.

[127] A. Elnaggar *et al.*, "ProtTrans: Towards Cracking the Language of Lifes Code Through Self-Supervised Deep Learning and High Performance Computing," *IEEE Transactions on Pattern Analysis and Machine Intelligence*, p. 1, 2021, doi: 10.1109/TPAMI.2021.3095381.

[128] A. Elnaggar *et al.*, "Ankh: Optimized Protein Language Model Unlocks General-Purpose Modelling." bioRxiv, Jan. 2023. doi: 10.1101/2023.01.16.524265.

[129] A. Schoenrock, F. Dehne, J. R. Green, A. Golshani, and S. Pitre, "MP-PIPE: A Massively Parallel Protein-Protein Interaction Prediction Engine," in *Proceedings of the International Conference on Supercomputing - ICS '11*, Tucson, Arizona, USA: ACM Press, 2011, p. 327. doi: 10.1145/1995896.1995946.

[130] J. Zhang *et al.*, "Computing the Human Interactome." bioRxiv, Oct. 2024. doi: 10.1101/2024.10.01.615885.

[131] M. Dayhoff, R. Schwartz, and B. Orcutt, "A Model of Evolutionary Change in Proteins," *Atlas of protein sequence and structure*, vol. 5, pp. 345–352, 1978.

[132] E. W. Bell, J. H. Schwartz, P. L. Freddolino, and Y. Zhang, "PEPPI: Whole-proteome Protein-protein Interaction Prediction through Structure and Sequence Similarity, Functional Association, and Machine Learning," *Journal of Molecular Biology*, vol. 434, no. 11, p. 167530, Jun. 2022, doi: 10.1016/j.jmb.2022.167530.

[133] A. Calderone, L. Licata, and G. Cesareni, "VirusMentha: A New Resource for Virus-Host Protein Interactions," *Nucleic Acids Research*, vol. 43, no. Database issue, pp. D588–592, Jan. 2015, doi: 10.1093/nar/gku830.

[134] L. Salwinski, C. S. Miller, A. J. Smith, F. K. Pettit, J. U. Bowie, and D. Eisenberg, "The Database of Interacting Proteins: 2004 Update," *Nucleic Acids Research*, vol. 32, no. Database issue, pp. D449–451, Jan. 2004, doi: 10.1093/nar/gkh086.

[135] J. Das and H. Yu, "HINT: High-quality Protein Interactomes and Their Applications in Understanding Human Disease," *BMC Systems Biology*, vol. 6, no. 1, p. 92, Jul. 2012, doi: 10.1186/1752-0509-6-92.

[136] S. Peri *et al.*, "Human Protein Reference Database as a Discovery Resource for Proteomics," *Nucleic Acids Research*, vol. 32, no. Database issue, pp. D497–D501, Jan. 2004, doi: 10.1093/nar/gkh070.

[137] R. Mosca, A. Céol, A. Stein, R. Olivella, and P. Aloy, "3did: A Catalog of Domain-Based Interactions of Known Three-Dimensional Structure," *Nucleic Acids Research*, vol. 42, no. D1, pp. D374–D379, Jan. 2014, doi: 10.1093/nar/gkt887.

[138] R. D. Finn, B. L. Miller, J. Clements, and A. Bateman, "iPfam: A Database of Protein Family and Domain Interactions Found in the Protein Data Bank," *Nucleic Acids Research*, vol. 42, no. D1, pp. D364–D373, Jan. 2014, doi: 10.1093/nar/gkt1210.

[139] K. Dick, F. Dehne, A. Golshani, and J. R. Green, "Positome: A Method for Improving Protein-Protein Interaction Quality and Prediction Accuracy," in *2017 IEEE Conference on Computational Intelligence in Bioinformatics and Computational Biology (CIBCB)*, Manchester, United Kingdom: IEEE, Aug. 2017, pp. 1–8. doi: 10.1109/CIBCB.2017.8058545.

[140] Y. Z. Zhou, Y. Gao, and Y. Y. Zheng, "Prediction of Protein-Protein Interactions Using Local Description of Amino Acid Sequence," *Advances in Computer Science and Education Applications*. Springer, Berlin, Heidelberg, pp. 254–262, 2011. doi: 10.1007/978-3-642-22456-0_37.

[141] M. G. Ammari, C. R. Gresham, F. M. McCarthy, and B. Nanduri, "HPIDB 2.0: A Curated Database for Host–Pathogen Interactions," *Database*, vol. 2016, Jan. 2016, doi: 10.1093/database/baw103.

[142] T. Guirimand, S. Delmotte, and V. Navratil, "VirHostNet 2.0: Surfing on the Web of Virus/Host Molecular Interactions Data," *Nucleic Acids Research*, vol. 43, no. Database issue, pp. D583–587, Jan. 2015, doi: 10.1093/nar/gku1121.

[143] S. Durmuş Tekir *et al.*, "PHISTO: Pathogen–Host Interaction Search Tool," *Bioinformatics*, vol. 29, no. 10, pp. 1357–1358, May 2013, doi: 10.1093/bioinformatics/btt137.

[144] D. E. Gordon *et al.*, "A SARS-CoV-2 Protein Interaction Map Reveals Targets for Drug Repurposing," *Nature*, vol. 583, no. 7816, pp. 459–468, Jul. 2020, doi: 10.1038/s41586-020-2286-9.

[145] J. Li *et al.*, "Virus-Host Interactome and Proteomic Survey Reveal Potential Virulence Factors Influencing SARS-CoV-2 Pathogenesis," *Med (New York, N.Y.)*, vol. 2, no. 1, pp. 99–112, Jan. 2021, doi: 10.1016/j.medj.2020.07.002.

[146] T. Bepler and B. Berger, "Learning Protein Sequence Embeddings Using Information from Structure," in *International Conference on Learning Representations*, arXiv, 2019. doi: 10.48550/arXiv.1902.08661.

[147] K. Chen, L. A. Kurgan, and J. Ruan, "Prediction of Flexible/Rigid Regions from Protein Sequences Using k-Spaced Amino Acid Pairs," *BMC Structural Biology*, vol. 7, no. 1, pp. 1–13, Dec. 2007, doi: 10.1186/1472-6807-7-25.

[148] X.-Y. Pan, Y.-N. Zhang, and H.-B. Shen, "Large-Scale Prediction of Human Protein-Protein Interactions from Amino Acid Sequence Based on Latent Topic Features," *Journal of Proteome Research*, vol. 9, no. 10, pp. 4992–5001, Oct. 2010, doi: 10.1021/pr100618t.

[149] Y. Feng, Q. Wang, and T. Wang, "Drug Target Protein-Protein Interaction Networks: A Systematic Perspective," *BioMed Research International*, vol. 2017, p. 1289259, 2017, doi: 10.1155/2017/1289259.

[150] J. M. Harrold, M. Ramanathan, and D. E. Mager, "Network-Based Approaches in Drug Discovery and Early Development," *Clinical Pharmacology and Therapeutics*, vol. 94, no. 6, pp. 651–658, Dec. 2013, doi: 10.1038/clpt.2013.176.

[151] Y.-A. Kim, S. Wuchty, and T. M. Przytycka, "Identifying Causal Genes and Dysregulated Pathways in Complex Diseases," *PLOS Computational Biology*, vol. 7, no. 3, p. e1001095, Mar. 2011, doi: 10.1371/journal.pcbi.1001095.

[152] M. A. Basar, M. F. Hosen, B. Kumar Paul, M. R. Hasan, S. M. Shamim, and T. Bhuyian, "Identification of Drug and Protein-Protein Interaction Network among Stress and Depression: A Bioinformatics Approach," *Informatics in Medicine Unlocked*, vol. 37, p. 101174, Jan. 2023, doi: 10.1016/j.imu.2023.101174.

[153] Q. Peng and N. J. Schork, "Utility of Network Integrity Methods in Therapeutic Target Identification," *Frontiers in Genetics*, vol. 5, p. 12, 2014, doi: 10.3389/fgene.2014.00012.

[154] F. E. Agamah et al., "Computational/in Silico Methods in Drug Target and Lead Prediction," *Briefings in Bioinformatics*, vol. 21, no. 5, pp. 1663–1675, Nov. 2019, doi: 10.1093/bib/bbz103.

[155] X. Zhang et al., "In Silico Methods for Identification of Potential Therapeutic Targets," *Interdisciplinary Sciences: Computational Life Sciences*, vol. 14, no. 2, pp. 285–310, Jun. 2022, doi: 10.1007/s12539-021-00491-y.

[156] L. Wang et al., "Therapeutic Peptides: Current Applications and Future Directions," *Signal Transduction and Targeted Therapy*, vol. 7, no. 1, p. 48, Feb. 2022, doi: 10.1038/s41392-022-00904-4.

[157] E. Rosson, F. Lux, L. David, Y. Godfrin, O. Tillement, and E. Thomas, "Focus on Therapeutic Peptides and Their Delivery," *International Journal of Pharmaceutics*, vol. 675, p. 125555, Apr. 2025, doi: 10.1016/j.ijpharm.2025.125555.

[158] W. Cabri et al., "Therapeutic Peptides Targeting PPI in Clinical Development: Overview, Mechanism of Action and Perspectives," *Frontiers in Molecular Biosciences*, vol. 8, Jun. 2021, doi: 10.3389/fmolb.2021.697586.

[159] I. Coin, M. Beyermann, and M. Bienert, "Solid-Phase Peptide Synthesis: From Standard Procedures to the Synthesis of Difficult Sequences," *Nature Protocols*, vol. 2, no. 12, pp. 3247–3256, Dec. 2007, doi: 10.1038/nprot.2007.454.

[160] A. Schoenrock et al., "Engineering Inhibitory Proteins with InSiPS: The in-Silico Protein Synthesizer," in *Proceedings of the International Conference for High Performance Computing, Networking, Storage and Analysis on - SC '15*, Austin, Texas: ACM Press, 2015, pp. 1–11. doi: 10.1145/2807591.2807630.

[161] D. Burnside et al., "In Silico Engineering of Synthetic Binding Proteins from Random Amino Acid Sequences," *iScience*, vol. 11, pp. 375–387, Jan. 2019, doi: 10.1016/j.isci.2018.11.038.

[162] M. Hajikarimlou et al., "A Computational Approach to Rapidly Design Peptides That Detect SARS-CoV-2 Surface Protein S," *NAR Genomics and Bioinformatics*, vol. 4, no. 3, p. lqac58, Jul. 2022, doi: 10.1093/nargab/lqac058.

[163] Y. Lei et al., "A Deep-Learning Framework for Multi-Level Peptide–Protein Interaction Prediction," *Nature Communications*, vol. 12, no. 1, p. 5465, Dec. 2021, doi: 10.1038/s41467-021-25772-4.

[164] K. Palepu et al., "Design of Peptide-Based Protein Degraders via Contrastive Deep Learning." Cold Spring Harbor Laboratory, May 2022. doi: 10.1101/2022.05.23.493169.

[165] T. Chen et al., "PepMLM: Target Sequence-Conditioned Generation of Peptide Binders via Masked Language Modeling," 2024.

[166] S. Bhat et al., "De Novo Design of Peptide Binders to Conformationally Diverse Targets with Contrastive Language Modeling," *Science Advances*, vol. 11, no. 4, p. eadr8638, Jan. 2025, doi: 10.1126/sciadv.adr8638.

[167] J. A. Ruffolo, J. J. Gray, and J. Sulam, "Deciphering Antibody Affinity Maturation with Language Models and Weakly Supervised Learning," no. arXiv:2112.07782. arXiv, Dec. 2021. doi: 10.48550/arXiv.2112.07782.

[168] J. Devlin, M.-W. Chang, K. Lee, and K. Toutanova, "BERT: Pre-training of Deep Bidirectional Transformers for Language Understanding," no. arXiv:1810.04805. arXiv, May 2019. doi: 10.48550/arXiv.1810.04805.

[169] J. Leem, L. S. Mitchell, J. H. R. Farmery, J. Barton, and J. D. Galson, "Deciphering the Language of Antibodies Using Self-Supervised Learning," *Patterns*, vol. 3, no. 7, p. 100513, Jul. 2022, doi: 10.1016/j.patter.2022.100513.

[170] Y. Liu *et al.*, "RoBERTa: A Robustly Optimized BERT Pretraining Approach," no. arXiv:1907.11692. arXiv, Jul. 2019. doi: 10.48550/arXiv.1907.11692.

[171] H. Kenlay, F. A. Dreyer, A. Kovaltsuk, D. Miketa, D. Pires, and C. M. Deane, "Large Scale Paired Antibody Language Models," *PLOS Computational Biology*, vol. 20, no. 12, p. e1012646, Dec. 2024, doi: 10.1371/journal.pcbi.1012646.

[172] C. Raffel *et al.*, "Exploring the Limits of Transfer Learning with a Unified Text-to-Text Transformer," no. arXiv:1910.10683. arXiv, 2019. doi: 10.48550/arXiv.1910.10683.

[173] T. H. Olsen, I. H. Moal, and C. M. Deane, "AbLang: An Antibody Language Model for Completing Antibody Sequences," *Bioinformatics Advances*, vol. 2, no. 1, p. vbac46, Jan. 2022, doi: 10.1093/bioadv/vbac046.

[174] R. W. Shuai, J. A. Ruffolo, and J. J. Gray, "IgLM: Infilling Language Modeling for Antibody Sequence Design," *Cell Systems*, vol. 14, no. 11, pp. 979–989, Nov. 2023, doi: 10.1016/j.cels.2023.10.001.

[175] A. Radford, J. Wu, R. Child, D. Luan, D. Amodei, and I. Sutskever, "Language Models Are Unsupervised Multitask Learners," 2019.

[176] A. Kovaltsuk, J. Leem, S. Kelm, J. Snowden, C. M. Deane, and K. Krawczyk, "Observed Antibody Space: A Resource for Data Mining Next-Generation Sequencing of Antibody Repertoires," *The Journal of Immunology*, vol. 201, no. 8, pp. 2502–2509, Oct. 2018, doi: 10.4049/jimmunol.1800708.

[177] B. L. Hie *et al.*, "Efficient Evolution of Human Antibodies from General Protein Language Models," *Nature Biotechnology*, vol. 42, no. 2, pp. 275–283, Feb. 2024, doi: 10.1038/s41587-023-01763-2.

[178] A. Rives *et al.*, "Biological Structure and Function Emerge from Scaling Unsupervised Learning to 250 Million Protein Sequences," *Proceedings of the National Academy of Sciences*, vol. 118, no. 15, p. e2016239118, Apr. 2021, doi: 10.1073/pnas.2016239118.

[179] J. Meier, R. Rao, R. Verkuil, J. Liu, T. Sercu, and A. Rives, "Language Models Enable Zero-Shot Prediction of the Effects of Mutations on Protein Function," in *35th Conference on Neural Information Processing Systems*, bioRxiv, Nov. 2021. doi: 10.1101/2021.07.09.450648.

[180] S. Boshar, E. Trop, B. P. de Almeida, L. Copoiu, and T. Pierrot, "Are Genomic Language Models All You Need? Exploring Genomic Language Models on Protein Downstream Tasks," *Bioinformatics*, vol. 40, no. 9, p. btae529, Sep. 2024, doi: 10.1093/bioinformatics/btae529.

[181] M. E. Consens, B. Li, A. R. Poetsch, and S. Gilbert, "Genomic Language Models Could Transform Medicine but Not Yet," *npj Digital Medicine*, vol. 8, no. 1, p. 212, Apr. 2025, doi: 10.1038/s41746-025-01603-4.

[182] S. Ali *et al.*, "Large Language Models in Genomics—A Perspective on Personalized Medicine," *Bioengineering*, vol. 12, no. 5, p. 440, Apr. 2025, doi: 10.3390/bioengineering12050440.

[183] Y. Li, G. Qiao, and G. Wang, "scKEPLM: Knowledge Enhanced Large-Scale Pre-Trained Language Model for Single-Cell Transcriptomics." bioRxiv, Jul. 2024. doi: 10.1101/2024.07.09.602633.

[184] Y. Zeng *et al.*, "CellFM: A Large-Scale Foundation Model Pre-Trained on Transcriptomics of 100 Million Human Cells," *Nature Communications*, vol. 16, no. 1, p. 4679, May 2025, doi: 10.1038/s41467-025-59926-5.

[185] K. Smietana, M. Siatkowski, and M. Møller, "Trends in Clinical Success Rates," *Nature Reviews Drug Discovery*, vol. 15, no. 6, pp. 379–380, Jun. 2016, doi: 10.1038/nrd.2016.85.

## Acknowledgments

The authors wish to thank the Natural Sciences and Engineering Research Council of Canada for its financial support.

## Author contributions

François Charih (FC) conducted the literature review and drafted the manuscript. Kyle K. Biggar (KKB) and James R. Green (JRG) acquired resources, reviewed and edited the manuscript. All authors read and approved of the manuscript.

## Potential conflicts of interest

The authors have no conflicts of interest to disclose.